\documentclass[pdflatex,sn-mathphys-num]{sn-jnl}

\usepackage{graphicx}%
\usepackage{multirow}%
\usepackage{amsmath,amssymb,amsfonts}%
\usepackage{amsthm}%
\usepackage{mathrsfs}%
\usepackage[title]{appendix}%
\usepackage{xcolor}%
\usepackage{textcomp}%
\usepackage{manyfoot}%
\usepackage{booktabs}%
\usepackage{listings}%
\usepackage{mdframed}

\theoremstyle{thmstyleone}%
\theoremstyle{thmstyletwo}%

\theoremstyle{thmstylethree}%

\begin{document}

\title[Article Title]{Why Do Pull Requests Go Silent? Uncovering the Barriers to Contribution Completion in Open-Source Code Review}


\author*{\fnm{Md Shamimur} \sur{Rahman}}\email{mdr614@usask.ca}

\author{\fnm{Farhana} \sur{Akter}}\email{pgf490@usask.ca}

\author{\fnm{Md Mustakim} \sur{Billah}}\email{mustakim.billah@usask.ca}
\author{\fnm{Zadia} \sur{Codabux}}\email{zadiacodabux@ieee.org}
\author{\fnm{Chanchal K.} \sur{Roy}}\email{croy@cs.usask.ca}

\affil{\orgdiv{Department of Computer Science}, \orgname{University of Saskatchewan}, \city{Saskatoon}, \state{Saskatchewan}, \country{Canada}}




\abstract{Pull requests (PRs) are central to pull-based software development, enabling distributed code review and collaborative contribution in open-source projects. However, many PRs become inactive before integration and are eventually abandoned or closed, resulting in wasted effort for both contributors and maintainers. Although prior studies have examined PR abandonment and review delays, less is known about the specific contribution types, discussion-level barriers, and post-stalling collaboration patterns associated with PR inactivity. This study investigates what types of PRs are most common among stalled contributions, why inactivity occurs from authors' and reviewers' perspectives, and how stalling is associated with subsequent contributor and reviewer engagement. We analyzed 14,234 stalled PRs and 164,562 review comments from 19 popular GitHub repositories that used stale-bot workflows. We developed an LLM-based voting classifier to categorize stalled PRs by contribution type and combined quantitative analysis with qualitative coding of general and inline review discussions. Our findings show that feature-enhancement and issue-fixing PRs constitute the largest share of stalled contributions, together accounting for over 77\% of classified stalled PRs. General review comments indicate that stalling is mainly associated with communication and coordination breakdowns, including missing interaction, delayed feedback, and unclear follow-up. Inline review comments further show that inactivity is not always caused by disengagement, as many PRs are blocked by technical and dependency-related issues such as failing checks, configuration problems, compatibility concerns, and environment mismatches. We also find that only 39.56\% of contributors returned to submit subsequent PRs, while reviewer re-engagement with the same contributors was approximately 21\%. These results suggest that PR inactivity is a socio-technical coordination problem involving communication, technical readiness, review ownership, and automation practices. Based on these findings, we provide recommendations for type-aware triage, clearer review feedback, explicit ownership of next actions, CI-related blocker management, and cause-aware stale-bot interventions.
}




\maketitle

\section{Introduction}\label{sec1}

Pull-based development, widely used in the open-source community \cite{gousios2014exploratory}, enables distributed contributions by having developers fork projects and submit code changes (e.g., new features, optimizations, bug-fixing, etc.) as Pull Requests (PRs) for merging into the main branch. These PRs are reviewed by project maintainers or expert developers, a process that not only improves software quality \cite{tao2015partitioning, bavota2015four, barnett2015helping} but also identifies code issues such as bugs, technical debt, and code smells \cite{wang2015comparative, bosu2016process, morales2015code, mcintosh2016empirical}. In addition, code review enhances design, testing, and security \cite{braz2022less, paixao2019impact, spadini2019test}, fosters team awareness \cite{bacchelli2013expectations}, facilitates knowledge transfer \cite{caulo2020knowledge}, and improves collaboration \cite{zampetti2022using}. It also drives important discussions on the relevance of code changes and implementation details, as reviewers seek clarifications and provide feedback while authors justify their approaches to facilitate patch acceptance. However, the process becomes wasteful or inefficient when contributions are left unfinished \cite{khatoonabadi2023wasted, khatoonabadi2023understanding, wessel2019should}. An industrial report\footnote{\url{https://codeclimate.com/blog/abandoned-pull-requests}} estimates that 8\% of PRs are wasted and never merged. These PRs are either rejected by maintainers or abandoned by their contributors, often representing valid contributions, however, they remain incomplete. Consequently, significant time and effort are wasted by both contributors preparing and submitting PRs and maintainers managing and reviewing them. Prior studies indicate that PRs are left unresolved either because contributors do not address maintainers’ comments or because maintainers fail to follow up on progress \cite{khatoonabadi2023wasted, li2021you, wang2019my}. Over time, these inactive PRs accumulate, cluttering the PR list and hindering the maintainers’ ability to manage and prioritize unresolved submissions \cite{gousios2015work, li2021you}.

Since popular open-source projects usually attract many contributors and receive a high volume of PRs daily, maintainers who are already juggling multiple tasks often find it challenging to manually track inactive PRs, monitor their progress, and close them when necessary \cite{khatoonabadi2023wasted, li2021you}. To alleviate this burden, the Stale Bot\footnote{\url{https://github.com/actions/stale}\label{stalebot}} was released in 2017 and has since been widely adopted by GitHub open-source projects. The bot automatically labels issues and PRs as ``Stale" after 60 days of inactivity, prompting contributors to take action or clarify priorities, and then closes them after an additional seven days of inactivity. However, the bot's effectiveness is subject to ongoing debate. Some community members have criticized the Stale Bot as being ``harmful\footnote{\url{https://drewdevault.com/2021/10/26/stalebot.html}}'' and a ``false economy\footnote{\url{https://blog.benwinding.com/github-stale-bots/index.html}}'', while others initially found it beneficial, they later began viewing it as problematic \cite{khatoonabadi2023understanding}. Additionally, studies have noted that the Stale Bot can contribute to noise and create friction for both authors and maintainers by generating fragmented information \cite{farah2022exploratory, rahman2022towards, wessel2021don}. 

Although integrating the Stale Bot helps reduce the manual burden by notifying contributors of inactivity, the reasons behind this inactivity have not been thoroughly explored, especially when PR authors have invested significant effort in making code changes and submitting them for review. Li et al. \cite{li2021you} examined 321 abandoned PRs and surveyed 710 open-source developers, identifying several high-level factors contributing to abandonment, including a lack of reviewer responsiveness and consensus, limited time and interest among PR authors, and process-related issues such as obsolescence and tedious review procedures. Similarly, Khatoonabadi et al. \cite{khatoonabadi2023wasted} analyzed a random sample of 354 abandoned PRs and found that most contributors abandoned their PRs without any explanation. Their study pinpointed ten major reasons for challenges, such as addressing reviewers’ comments, insufficient reviews from maintainers, and difficulties resolving Continuous Integration (CI) failures, which occur most frequently and many of which overlap with the findings of Li et al. \cite{li2021you}. However, these studies leave several critical gaps unaddressed. It remains unclear which specific types of PRs, whether categorized by size, complexity, or purpose, are most prone to abandonment. While both studies acknowledge the challenge of addressing reviewer comments as a potential reason for abandonment, they do not investigate the nature of these comments (for example, whether they are overly technical, vague, or conflicting) or provide detailed perspectives from PR authors on the difficulties they encounter. Furthermore, the fine-grained reasons for abandonment, from the viewpoints of both PR authors and reviewers, are insufficiently explored. The consequences of abandonment are also unexplored. For example, if reviewers are primarily responsible for PR stalling, how does this impact authors' motivation and their continued participation in open-source projects, especially when their previous contributions were disregarded? In addition, neither study considers how abandonment affects the behavior of reviewers and contributors over time. Finally, there is a lack of actionable recommendations to help reduce PR abandonment, leaving open-source communities without clear guidance.  

In this study, we investigate 14,234 stalled PRs and 164,562 code review comments collected from 19 popular open source GitHub projects, which were also examined in the previous study by Khatoonabadi et al. \cite{khatoonabadi2023understanding}. Moreover, we adopt a Large Language Model (LLM) based classification technique to categorize the types of stalled PRs, aiming to identify which types of PRs, such as bug fixes, feature enhancements, or code optimizations, are more likely to become inactive or abandoned by PR authors. We then perform a detailed analysis of the reasons behind inactivity, considering both general and inline review comments. To understand the impact on individuals, we separately explore cases where either PR authors or reviewers contributed to the inactivity. Finally, we present several practical recommendations to help prevent such situations, where both authors and reviewers may unintentionally waste time and effort, potentially affecting the open-source contribution process. Our key contributions include the following: 

\begin{itemize}

    \item \textbf{Automatic PR type classification:} Developed an LLM-based voting approach to classify stalled PRs by contribution type and characterize which types of contributions are most frequently represented among inactive PRs (e.g., bug fixes, enhancements).
    
    \item \textbf{Fine-grained analysis of inactivity causes:} Performed a detailed investigation of the rationale behind PR inactivity by examining both general and inline review comments.
    
    \item \textbf{Responsibility attribution and impact analysis:} Identified possible impacts in cases where PR authors were responsible for the inactivity, and vice versa, where reviewers contributed to the stalling of PRs.
    
    \item \textbf{Actionable recommendations:} Proposed practical suggestions to help avoid PR inactivity, reduce wasted effort, and improve the efficiency of open source collaboration. Additionally, we shared the collected dataset, scripts of experiments, and associated results in the replication package\footnote{\url{https://doi.org/10.5281/zenodo.20599077}\label{replication package}}.
\end{itemize}

\section{Background and Related Work}
PR abandonment, where a contribution stalls and is never brought to completion, has become a persistent challenge in pull-based development. This section synthesizes prior research relevant to how inactivity is handled in pull-based contribution workflows and how automation tools may shape contribution trajectories.

\subsection{Abandoned PR}
In pull-based development workflows, contributors typically begin by forking or cloning an upstream repository to create a local, isolated environment for implementing and testing code changes. Once the changes are ready, they are pushed to a feature branch in the contributor’s fork, followed by the submission of a PR to the original repository. This action notifies project integrators, who initiate the review process \cite{tsay2014let, yu2015wait}. PRs commonly undergo multiple rounds of evaluation, during which integrators assess both the technical quality of the proposed changes and their compliance with project-specific conventions. If deficiencies such as coding standard violations, missing test cases, or insufficient documentation are identified, contributors are asked to make revisions and resubmit their work \cite{tsay2014let, yu2015wait}. This iterative process continues until consensus is reached regarding the adequacy of the contribution \cite{lenarduzzi2021does}.
Acceptance decisions are influenced not only by technical indicators such as code complexity, the number of lines modified, or test coverage, but also by social and behavioral factors, including contributor reputation, previous interactions, and the clarity of communication within the discussion thread \cite{tsay2014influence}. However, empirical studies have shown that contributors sometimes disengage before addressing requested changes, leading to PR abandonment \cite{khatoonabadi2023wasted, li2021you}. The majority of abandonment cases are attributed to contributors losing interest or lacking sufficient time, though additional qualitative findings highlight other contributing factors such as the complexity of required revisions and challenges in achieving agreement among reviewers \cite{khatoonabadi2023wasted}.
On the other hand, delays in receiving feedback from reviewers or integrators can also lead to inactivity. Yu et al. \cite{yu2015wait} observed significant variability in PR evaluation latency across projects, with integrator workload and project governance practices playing a key role. Moreover, recent research has identified several factors that impact review latency, including the size of the PRs (e.g., number of commits or files changed), the depth of review required (e.g., volume of review comments), and reviewer availability (e.g., time-to-first-response) \cite{zhang2022pull, li2021you, yu2015wait}. These factors have been shown to correlate with higher abandonment rates, particularly when feedback is delayed beyond the contributor’s period of engagement or availability. 

\subsection{Stale Bot Adoption and Impact}
Several studies explored automation tools like the Stale bot\footref{stalebot} for managing issues and PRs in open-source projects. M. Wessel et al. \cite{wessel2019should} examined the adoption and configuration of the Stale bot, which automatically labels and closes inactive or abandoned issues and PRs in GitHub repositories. They also noted that bug report issues and PRs awaiting further input were usually exempt from stalling. However, their study only looked at how the bot is set up and did not provide any insight into how it affects project management or community involvement, leaving the broader effects of using bots in automation. Existing studies \cite{rahman2022towards, wessel2020inconvenient, wessel2021don} reported that the Stale bot can generate noise and cause friction for both authors and maintainers by producing fragmented information. More recently, Khatoonabadi et al. \cite{khatoonabadi2023understanding} examined the effects of the Stale bot across 20 prominent open-source projects. Their findings suggest that while the bot facilitates the management of outdated PRs and accelerates the review workflow, its use may inadvertently reduce contributor engagement over time.

\subsection{Factors Behind Evaluation of PR}

Recent research has thoroughly examined how both technical and social factors influence the acceptance of PRs and the time it takes to review them. Gousios et al. \cite{gousios2014exploratory, gousios2015work} showed that factors such as touching recently modified code, adherence to project style and architecture, code quality, and test coverage significantly affect merge decisions and processing time, while technical reasons account for only a small minority of PR rejections. Tsay et al. \cite{tsay2014influence} expanded this view by splitting determinants into social and technical categories, finding that PRs with extensive discussions are less likely to be accepted and that mature projects tend to adopt more conservative acceptance practices. Soares et al. \cite{soares2015acceptance} identified additional influences such as programming language, commit counts, files added, contributor status (external vs. internal), and first‑time contributor status on both merge decisions and latency. Yu et al. \cite{yu2015wait, yu2016determinants} highlighted the impact of PR size, first‑response delay, CI pipeline availability, and contributor trustworthiness on review dynamics, and Kononenko et al. \cite{kononenko2018studying} further linked review time and merge outcomes to factors such as the number of participants in discussions, the experience of contributors, their affiliations, the complexity of PR descriptions, and whether the PRs could be easily reverted. An industrial study by Pinto et al. \cite{pinto2018gets} confirmed that external contributors face higher rejection rates and longer wait times than employees, Zou et al. \cite{zou2019does} found that style‑violating PRs suffer more rejections and delays, and Lenarduzzi et al. \cite{lenarduzzi2021does} suggested that the reputation of maintainers and the importance of features might have a greater impact on acceptance decisions than the quality of the code itself.

Researchers have also investigated how demographic and personal characteristics relate to the outcomes of PRs. Terrell et al. \cite{terrell2017gender} identified gender-based disparities, particularly noting that external contributors whose gender could be inferred experienced lower acceptance rates. Similarly, Rastogi et al. \cite{rastogi2016biases, rastogi2018relationship} and Nadri et al. \cite{nadri2021insights, nadri2021relationship} highlighted biases linked to geographic proximity and perceived racial identity, with contributors from the same region or race as maintainers more likely to have their PRs accepted. Furtado et al. \cite{furtado2020successful} further found that developers from countries with lower human development indices tend to submit fewer PRs and face higher rejection rates. In addition to these demographic factors, individual personality traits also influence PR acceptance. Iyer et al. \cite{iyer2019effects, deyoung2007between} reported that contributors who score high in openness and conscientiousness, despite being low in extroversion, tend to have higher acceptance rates, while integrators who exhibit conscientiousness, extroversion, and neuroticism are more likely to approve PRs. Moreover, an important factor affecting PR decisions is the risk of duplication. This can waste the efforts of both contributors and reviewers, potentially leading to the rejection of PRs \cite{li2020redundancy, li2021detecting}.

Prior research has largely concentrated on bot configuration and has surfaced only broad, high‑level reasons for why individual contributors disengage. These studies neither distinguish among different PR types nor delve into the specific challenges that authors and reviewers encounter during the review process. In addition, they have overlooked the longer‑term effects of PR abandonment on contributor and maintainer motivation, engagement patterns, and collaboration behavior. This study addresses these gaps by: (i) developing a detailed taxonomy of PR abandonment, which encompasses technical, communication, and personal constraints from both the author’s and reviewer’s perspectives; (ii) empirically validating these causes across various project domains to demonstrate how abandonment dynamics impact and change based on PR characteristics; and (iii) translating our findings into practical recommendations for tooling enhancements that open-source communities can implement to support participation, improve handover practices, and reduce the human cost associated with stalled contributions.

\section{Methodology}
Our methodology, outlining the approach and processes undertaken to investigate and address the research questions, is as follows.

\subsection{Research Questions}
To thoroughly investigate stalled PRs that may lead to dissatisfaction among both PR authors and reviewers, we employed a mixed-methods approach combining qualitative and quantitative analyses. Our study examines the prevalence of different PR types, uncovers potential reasons for inactivity, and analyzes how PR inactivity may influence authors’ subsequent contribution behavior. Based on our findings, we propose actionable recommendations to reduce wasted effort and highlight the potential role of automated tools in mitigating such issues. Our analysis aims to address the following three Research Questions (RQs):

\setlength{\leftmargini}{0.9em}
\begin{itemize}
    \item \textbf{RQ$_1$: What types of PRs are most common among stalled PRs?} This RQ focuses on using an LLM-based classifier approach to identify the prevalence of stalling across different PR types (e.g., bug fixes, enhancements, and refactoring).

    \item \textbf{RQ$_2$: What factors contribute to PR inactivity in open-source projects?} Here, we aim to uncover detailed and nuanced causes of stalling by analyzing general and inline review comments.
    
    \item \textbf{RQ$_3$: Where does the next required action reside in stalled PRs, and how is stalling associated with subsequent collaboration?} This explores the attribution of inactivity, distinguishing between author-driven and reviewer-driven causes, and considers how each influences contributor engagement and satisfaction.
 \end{itemize}   

\subsection{Data Collection}
Several existing studies have explored PR abandonment using different open-source projects. However, our focus is on investigating not just the reasons behind PR abandonment but also its effects on contributors and maintainers in projects that actively use the Stale bot to manage inactivity. To ensure the reliability and adequacy of our dataset,  we based our project selection on the 20 open-source projects referenced in the prior study by Khatoonabadi et al. \cite{khatoonabadi2023understanding}, which were previously examined for Stale bot interventions in managing PRs\footnote{\url{https://doi.org/10.5281/zenodo.7978381}}. During data collection, we discovered that one project was no longer available on GitHub, so our analysis proceeded with the remaining 19 projects. These projects comprise a diverse set of active, well-maintained open-source repositories that have consistently used the Stale bot to mark and close inactive PRs. Using the GitHub API\footnote{\url{https://docs.github.com/en/rest?apiVersion=2022-11-28}}, we collected PRs' information that were marked as stalled by the Stale bot during their lifecycle. For each PR, we retrieved metadata including titles, descriptions, general comments, and inline review comments. The statistics of the resulting dataset are presented in Table \ref{table 1}.  

\begin{table}[htbp]
\caption{Dataset Statistics}
\centering
\label{table 1}
\begin{tabular}{@{}lr@{}}
\toprule
\textbf{Steps}                                             & \textbf{\#Quantity} \\ \midrule
Number of projects                                & 19         \\
Number of total PRs                               &  1,210,878          \\
Number of stalled PRs                             &  14,234          \\
Number of general review comments in stalled PRs  &  1,14,184        \\ 
Number of inline review comments in stalled PRs   &  50,378          \\
Number of average LOC changes in each stalled PRs &  $\sim 293$  \\\midrule      
\end{tabular}
\end{table}

\subsection{Stalled PR Classification}
Harbaoui et al. \cite{harbaoui2024impact} draft PRs into 16 distinct types. Building on their classification, we grouped these categories into seven broader themes: \textit{Feature Enhancement and Upgrading}, \textit{Fixing Issues}, \textit{Code Optimization}, \textit{Database Optimization}, \textit{UI Improvement and Enhancement}, \textit{Version and Dependency Management}, and \textit{Protocol Improvements}. This consolidation aims to offer a more precise and comprehensive perspective on the common reasons behind stalled PRs. To efficiently categorize PRs across our collected dataset, we needed an automated and reliable labeling method. Manually labeling over 14K PRs was impractical. To overcome this challenge, we adopted an LLM-based classification strategy, following approaches used in recent studies \cite{wang2025can, ahmed2025can, hassani2025empirical}. Specifically, we employed three open-source LLMs: Llama 3.3:70B\footnote{\url{https://ollama.com/library/llama3.3}}, Qwen2:7B-Instruct\footnote{\url{https://ollama.com/library/qwen2}}, and DeepSeek R1:14B\footnote{\url{https://ollama.com/library/deepseek-r1}}, selected for their strong inference capabilities and relevance in prior research \cite{li2024development, guo2025deepseek, hou2024large}.  

\begin{figure}[htbp]
\centerline{\includegraphics[width=1\linewidth]{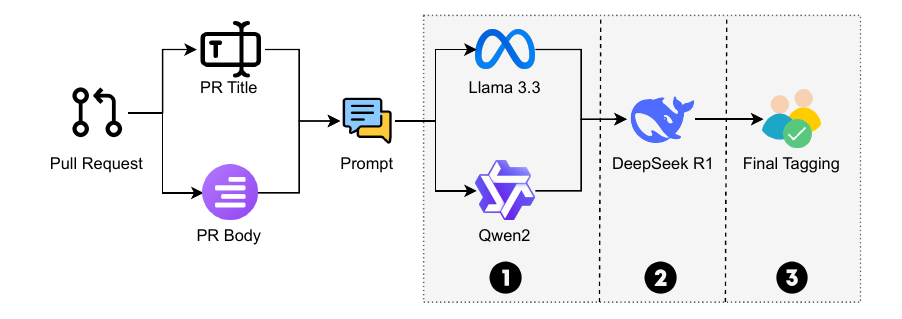}}
\caption{PR classification using LLM-based voting: \textcircled{\small{1}} Llama and Qwen independently label each PR; \textcircled{\small{2}} DeepSeek resolves disagreements between Llama and Qwen; \textcircled{\small{3}} remaining conflicts are manually labeled.}
\label{fig:prClassify}
\end{figure}

Each PR includes a concise title and a detailed description (i.e., PR body), which often includes module names, issue references, and implementation rationale. Since titles alone often lack sufficient detail to convey the functionality or full intent of the PR, we used both the title and the description to classify PRs into the seven predefined categories. To reduce noise and improve input quality, we first cleaned the PR descriptions by removing project-specific template content, following the method by Liu et al. \cite{liu2019automatic}. As part of preprocessing, we removed URLs, hashtags, user signatures, and email addresses from both the titles and descriptions. After preprocessing, we retained only the PRs with non-empty titles and bodies, resulting in a final dataset of 12,817 PRs. We then designed an optimized prompt, guided by OpenAI’s prompt engineering principles\footnote{\url{https://platform.openai.com/docs/guides/prompt-engineering}}, to support LLM-based classification. Our classification process follows a voting-based strategy \cite{dietterich2000ensemble, snow2008cheap}, illustrated in Figure~\ref{fig:prClassify}. Each PR was first classified independently by Llama 3.3 and Qwen2. If their outputs matched, the label was accepted. In cases of disagreement, we introduced a third label from DeepSeek R1. The final category was determined by majority vote. In the event of a three-way disagreement, one coder with more than 10 years of software development experience manually reviewed the PR title and description and assigned the final label using the predefined category definitions. To reduce individual bias, these manually resolved cases were later cross-checked by the other two coders during the verification stage.

\subsection{Comment Analysis}\label{comment analysis}
When a PR is created, collaborators can provide feedback through two main types of comments: inline review comments, which target specific code changes (i.e., DiffHunk), and general review comments, which address the overall code modifications\footnote{\url{https://tinyurl.com/2rx5efz6}}. To investigate the nuanced factors contributing to PR stalling, we analyzed both comment types separately across 14,234 PRs. We made this distinction because inline comments are typically tied to code change requests, often instructing contributors to revise or improve specific portions of the code, while general comments tend to focus on broader aspects of the PR, such as design decisions, testing coverage, or the overall scope of the changes. This separation allows us to better capture the distinct review dynamics that may contribute to a PR stalling.  Furthermore, to capture the full review context for inline change requests, we merged all inline review comments in each PR into a single consolidated thread. This process resulted in a total of 6,333 inline review comment threads, as not every PR contained them. To reduce the time and effort of manual comment analysis, we used a random sampling approach. From 14,234 PRs, we selected 375 general review comment threads and 363 inline review comment threads, with a 95\% confidence level and 5\% margin of error. To minimize bias, we created four independent, non-overlapping samples for each comment type, totaling 1,500 general reviews and 1,452 inline reviews. 

To investigate the underlying factors contributing to PR stalling, we conducted an in-depth thematic analysis \cite{cruzes2011recommended} for both comment types. This qualitative inquiry sought to uncover recurring themes, behavioral patterns, and contextual elements that elucidate delays in the PR review process. Three individuals with more than 10, nine, and six years of software development experience, respectively, participated in the coding process, which followed a systematic and iterative methodology. Initially, each rater independently performed open coding on a randomly selected subset comprising 50 general and 50 inline review comment threads, drawn from the sampled data. This initial coding phase yielded a diverse set of stalling-related concepts, which were then collaboratively reviewed. Through iterative consensus discussions, the preliminary codes were refined and organized into a structured codebook containing higher-order thematic categories.

To maintain consistency and minimize subjective bias during the annotation process, the rest of the comment threads were independently double-coded by at least two raters using the finalized coding schema. Inter-rater reliability was assessed using Cohen’s Kappa~\cite{cohenkappa}, resulting in an average $\kappa$ value of 0.74 across all threads, which indicates substantial agreement. In instances of disagreement, the third rater served as a mediator to adjudicate and finalize the labels based on majority consensus. Thus, our systematic process ensured that the final taxonomy of reasons for PR stalling was grounded in empirical data and expert consensus. These categories, along with their associated findings, are presented in the subsequent results section.

\subsection{Post-Stale Engagement}
To evaluate the post-engagement patterns of contributors and reviewers in the context of stalled PRs, we began by identifying the unique authors and reviewers associated with each stalled PR in a given project. For each contributor, we examined their next 10 PRs (if any) submitted to the same repository following the stalling event. This analysis yielded two possible outcomes: (1) the contributor submitted no subsequent PRs, which may indicate that the stalled PR experience likely discouraged continued participation, or (2) the contributor remained active, suggesting that the stalling did not significantly deter future engagement.

To assess reviewer behavior, we analyzed whether the same reviewers who were involved in a contributor’s stalled PR also reviewed any of their subsequent 10 PRs. A high recurrence of reviewer-author pairs would imply that reviewers maintain objectivity and prioritize technical content over contributor identity. In contrast, a low recurrence rate could suggest hesitancy or selective disengagement from prior interactions with contributors, potentially reflecting interpersonal or implicit bias dynamics \cite{soares2018factors, rahman2023integrating}.

\begin{table}[htbp]
\centering
\caption{Distribution of PRs across consolidated categories}
\label{tab:category-distribution}
\begin{tabular}{lrr}
\toprule
\textbf{Category} & \textbf{\# PRs} & \textbf{Percentage} \\
\midrule
Feature Enhancement and Upgrading & 6,927 & 54.04\% \\
Fixing Issues                     & 3,029 & 23.63\% \\
Version and Dependency Management & 1,162 & 9.07\%  \\
Code Optimization                  & 1,086 & 8.47\%  \\
UI Improvement and Enhancement     & 277   & 2.16\%  \\
Protocol Improvements              & 182   & 1.42\%  \\
Database Optimization              & 151   & 1.18\%  \\
\bottomrule
\end{tabular}
\end{table}

\section{Results} \label{sec:Results}

\subsection{RQ$_1$: PR Types Prone to Stalling}
Of the 12,817 PRs, Llama3.3 and Qwen2 agreed on the classification of 10,376 PRs (81\%). Among the remaining 2,441 PRs (19\%) where they disagreed, DeepSeek R1 resolved 2,398 cases, agreeing with Llama3.3 on 1,672 and with Qwen2 on 726. This left only 43 PRs (0.3\% of the total) for the first author to manually review and classify.

Table \ref{tab:category-distribution} shows the distribution of PRs across the categories. Feature Enhancement and Upgrading emerged as the most frequent category, accounting for 54.04\% of the PRs, followed by Fixing Issues (23.63\%), and Version and Dependency Management (9.07\%). Code Optimization also contributed notably with 8.47\% of the PRs. Less frequent categories included UI Improvement and Enhancement (2.16\%), Protocol Improvements (1.42\%), and Database Optimization (1.18\%).

To test whether the category distribution significantly deviates from uniformity, we conducted a Chi-square goodness-of-fit~\cite{rolke2021chi} test. The results showed a statistically significant deviation ($\chi^2 = 19,357.79$, $df = 6$, $p < 0.001$), indicating that PRs are not evenly distributed across categories. This suggests that certain types of PRs, particularly those related to Feature Enhancement and Upgrading and Fixing Issues, are more frequently associated with stalled contributions.

\begin{mdframed}[
    linecolor=black!60,
    linewidth=1.5pt,
    backgroundcolor=yellow!8,
    innertopmargin=6pt,
    innerbottommargin=6pt
]
\noindent\textbf{\underline{RQ1 Findings:}} \textbf{Feature Enhancement and Upgrading (54.04\%)} and \textbf{Fixing Issues (23.63\%)} are the most common contribution types among stalled PRs, together accounting for over 77\% of classified stalled cases. Less frequent categories include \textbf{Protocol Improvements (1.42\%)} and \textbf{Database Optimization (1.18\%)}. A Chi-square test confirmed that the distribution of stalled PRs across contribution categories is significantly non-uniform ($p < 0.001$), showing that stalled contributions are concentrated in a small number of PR types.
\end{mdframed}

\subsection{RQ$_2$: Rationale for PR inactivity}

To understand \textit{why} PR stalls, we conducted a systematic qualitative analysis of both general and inline review comments. Through open coding and iterative categorization (as discussed in Section \ref{comment analysis}), we identified six high-level categories comprising 22 subcategories from general review comments, and five high-level categories comprising 17 subcategories from inline review comments. Table~\ref{table 2} presents the taxonomy derived from general review comments, while Table~\ref{table 4} summarizes findings from inline review comments. The two analyses are conducted separately because general and inline review comments serve different purposes in the review process. General review comments usually capture PR-level assessments, such as the overall acceptability, readiness, or direction of a contribution~\cite{bacchelli2013expectations}. Inline comments, by contrast, are attached to specific code locations and therefore expose more localized concerns about implementation choices, code quality, testing, and maintainability~\cite{mcintosh2016empirical, bosu2016process}.

\begin{table*}[htbp]
\small
\caption{Common reasons behind stalled PRs from general review comments.}
\label{table 2}
\resizebox{\textwidth}{!}{%
\begin{tabular}{@{}p{4cm}p{3cm}p{11cm}@{}}
\toprule
\textbf{Category}                              & \textbf{Subcategory}                               & \textbf{Description}                                                                                   \\ \midrule
Collaboration \& Availability Issues (76.60\%) & Author Unresponsive (15.4\%)                        & Author has not replied to reviewer comments or feedback.                                       \\\cmidrule{2-3}
                                      & Reviewer Unresponsive (14.93\%)                      & Reviewer has not responded to author's updates or queries.                                     \\\cmidrule{2-3}
                                      & No Interaction (17.80\%)                             & No conversation between author and reviewer.                                                  \\\cmidrule{2-3}
                                      & Author Unavailable (1\%)                         & Author is unavailable or unable to make required changes or updates to the PR.                \\\cmidrule{2-3}
                                      & Reviewer Unavailable (0.87\%)                       & Reviewer is unavailable to review or provide feedback on the PR.                              \\\cmidrule{2-3}
                                      & No Follow-Up (7.47\%)                              & No follow-up from either author or reviewer after initial review or comment.                  \\\cmidrule{2-3}
                                      & No Progress on Issues (5.33\%)                      & No meaningful progress is made on the PR from either side.                                    \\\cmidrule{2-3}
                                      & Missing Feedback (11.93\%)                           & Reviewer has not provided actionable feedback after initial review.                           \\\cmidrule{2-3}
                                      & Unresolved Comments (1.87\%)                        & Issues or comments raised during the review are left unresolved.                              \\\cmidrule{1-3}
Reviewer Initiated Actions (5.13\%)     & PR Rejection (0.13\%)                               & Reviewer explicitly rejects the proposed changes in the PR.                                   \\\cmidrule{2-3}
                                      & Change Request (2.20\%)                            & Reviewer requests changes to the PR before approval.                                          \\\cmidrule{2-3}
                                      & Redirection (2.80\%)                                & Reviewer redirects the PR to another reviewer or maintainer for further handling.             \\\midrule
Conflict of Perspectives  (2.60\%)           & Disagreement (1.67\%)                               & Author and reviewer disagree on the approach or solution in the PR.                           \\\cmidrule{2-3}
                                      & Diverging Opinions( 0.93\%)                         & Author and reviewer hold fundamentally different views, making common ground difficult.           \\\cmidrule{1-3}
Technical \& Testing Issues  (7.33\%)        & Unresolved Test Failures (3.33\%)                   & Test failures remain unresolved, blocking the PR from moving forward.                         \\\cmidrule{2-3}
                                      & Technical Issues (3.80\%)                           & Technical problems in the PR, such as bugs or incompatibilities, preventing progress.         \\\cmidrule{2-3}
                                      & Non-Reproducible Issues (0.20\%)                    & Issues reported in the PR cannot be consistently reproduced or observed.                      \\\cmidrule{1-3}
Policy \& Procedural Issues (3.87\%)         & Contributor License Agreement (CLA) Issues (2.67\%) & PR cannot proceed due to missing or unprocessed CLA.                          \\\cmidrule{2-3}
                                      & Administrative Issue (1.13\%)                       & Procedural or governance-related issues are preventing the PR from being merged.              \\\cmidrule{2-3}
                                      & Redundant PR (0.07\%)                               & Multiple PRs with the same or similar changes, causing redundancy.                            \\\cmidrule{1-3}
External Constraints \& Limitations (4.47\%)  & External Factors (0.67\%)                           & External circumstances (e.g., third-party delays, infrastructure) prevent progress on the PR. \\\cmidrule{2-3}
                                      & External Dependency (3.80\%)                        & PR is blocked due to dependencies on external systems, libraries, or services.                \\ \bottomrule
\end{tabular}%
}
\end{table*}

\begin{table*}[htbp]
\small
\caption{Reasons for PR stalling identified from inline review comments.}
\label{table 4}
\resizebox{\textwidth}{!}{%
\begin{tabular}{@{}p{4cm}p{4cm}p{10cm}@{}}
\toprule
\textbf{Category}                                     & \textbf{Subcategory}                             & \textbf{Description}                                                                                                                                   \\ \midrule
Review Process \& Communication Issues (47.31\%)       & Unresolved Discussion (18.39\%)                   & Delays caused by ongoing clarifications, unclear direction, or feedback that has not been fully addressed.                                     \\\cmidrule{2-3}
                                             & Communication Delays (8.75\%)                    & Lack of response from reviewers/authors, waiting on maintainers or external confirmations.                                                    \\\cmidrule{2-3}
                                             & Final Approval Pending (14.39\%)                  & Delays caused by resolved issues awaiting final approval or completion of the review process.                                                 \\\cmidrule{2-3}
                                             & Review Process Challenges (5.79\%)               & Delays caused by multiple review iterations, limited reviewer availability, shifting priorities, large PRs, or reviewer fatigue               \\\cmidrule{1-3}
Technical \& Dependency Issues (22.45\%)               & Testing \& Environment Failures (4.06\%)         & Challenges with tests not passing or missing necessary tests, along with unresolved environment setup issues.                                 \\\cmidrule{2-3}
                                             & Dependency \& Configuration Problems (8.95\%)    & Problems arising from build configuration issues, missing dependencies, or gaps in CI processes.                                              \\\cmidrule{2-3}
                                             & Versioning \& Compatibility Issues (2.48\%)      & Concerns with compatibility between versions or unresolved versioning issues, including breaking changes.                                     \\\cmidrule{2-3}
                                             & Security \& Compliance Issues (1.24\%)           & Security vulnerabilities or compliance concerns such as licensing issues or problems with dependencies.                                       \\\cmidrule{2-3}
                                             & General Technical Issues (5.72\%)                & Broad technical problems, such as build failures, performance issues, or uncertainties related to the code.                                   \\\cmidrule{1-3}
Code Quality Issues (17.49\%)                          & Code Structure \& Optimization Problems (4.96\%) & Concerns regarding the structure of the code, optimization opportunities, refactoring needs, and enhancing readability or maintainability.    \\\cmidrule{2-3}
                                             & Styling \& Formatting Issues (7.71\%)            & Problems related to inconsistent code style, formatting errors, naming inconsistencies, or unresolved typos.                                  \\\cmidrule{2-3}
                                             & Documentation Gaps \& Inconsistencies (4.82\%)   & Issues with incomplete or inconsistent documentation, and unresolved naming or localization problems.                                         \\\cmidrule{1-3}
Implementation \& Decision-Making Challenges (9.92\%) & Design \& Architecture Conflicts (6.06\%)        & Disagreements over design choices, unresolved trade-offs, or pending decisions on the project’s architecture.                                 \\\cmidrule{2-3}
                                             & Performance \& Efficiency Issues (0.69\%)        & Issues related to system performance, including bottlenecks, inefficiencies, and suggestions for improving execution speed or resource usage. \\\cmidrule{2-3}
                                             & Implementation Challenges (3.17\%)               & Issues related to incomplete implementation, progress being blocked, or pending updates from the author that prevent moving forward           \\\cmidrule{1-3}

Merging \& Review Delays (2.82\%)                     & Conflicts \& Merge Issues (0.96\%)               & Problems arising from merge conflicts, redundant pull requests, or merges being blocked due to reviewer requests.                             \\\cmidrule{2-3}
                                             & Iterative Review \& Refinements Delay (1.86\%)   & Delays caused by repeated revisions, pending code changes, or an ongoing back-and-forth during the review process.                            \\ \bottomrule
\end{tabular}%
}

\end{table*}

\subsubsection{General Review Comment}\label{general review}
General review comments reveal that PR inactivity is primarily associated with breakdowns in coordination, responsiveness, and review ownership. To gain a deeper understanding of the six identified reasons, we further divided them into 22 more specific subcategories. 
\\

\textbf{Collaboration \& Availability Issues.} Among the six categories identified from these comments, \textit{Collaboration and Availability Issues} dominate, accounting for 76.60\% of the observed stalling reasons, which encompasses a broad spectrum of breakdowns in the human-centric aspects of the review process. This category includes nine subcategories and captures situations where the review process loses momentum because one or more participants do not provide the next action required to move the PR forward.

The most frequent subcategory is \textit{No Interaction} (17.80\%), where the PR shows little or no substantive exchange between the author and reviewer. This suggests that some PRs become inactive before meaningful review collaboration is established. \textit{Author Unresponsive} (15.40\%) and \textit{Reviewer Unresponsive} (14.93\%) further show that inactivity can originate from either side of the review process. In some cases, reviewers provide feedback, but authors do not return to address it. In others, authors submit updates or seek clarification, but reviewers do not provide the next review action. Although these patterns differ in source, they converge on the same outcome: the PR remains open without progressing.

Other subcategories refine this coordination problem. \textit{Missing Feedback} (11.93\%) captures cases where reviewers do not provide actionable feedback after an initial interaction. \textit{No Follow-Up} (7.47\%) and \textit{No Progress on Issues} (5.33\%) indicate that even when communication begins, the absence of sustained engagement can prevent the PR from reaching resolution. Less frequent cases, such as \textit{Author Unavailable} (1.00\%) and \textit{Reviewer Unavailable} (0.87\%), show that practical constraints such as temporary absence or limited availability can also interrupt the review lifecycle. For instance, in one Apache Beam PR\footnote{\url{https://github.com/apache/beam/pull/6004}}, the reviewer noted that they were on holiday and planned to resume the review after returning: \textit{“... I am currently on holidays but plan to tackle this after my return (next week) ...”} This example illustrates how even temporary reviewer unavailability can delay progress when the PR depends on a specific reviewer or maintainer.
\\


\textbf{Reviewer Initiated Actions.} This category captures cases where reviewer actions themselves contribute to stalling. Unlike collaboration and availability issues, which are often characterized by missing or delayed action, reviewer-initiated actions involve explicit review decisions that alter the trajectory of the PR. The category includes Change Request (2.20\%), Redirection (2.80\%), and PR Rejection (0.13\%). A \textit{Change Request} pauses the PR until the author addresses the requested modifications. \textit{Redirection} transfers responsibility to another reviewer or maintainer, which may introduce handoff delays. \textit{PR Rejection} reflects cases where the reviewer determines that the proposed contribution should not proceed in its current form. Although these actions are legitimate parts of the review process, they can still contribute to inactivity when the author does not revise the PR, when the redirected reviewer does not respond, or when the rejection leaves the PR open without further action.

For example, in a gRPC PR\footnote{\url{https://github.com/grpc/grpc/pull/23050}}, the reviewer rejected the proposed fix by explaining that the submitted change addressed the wrong build system: \textit{“This is probably wrong—you’re saying that the Homebrew installation ... is broken, but you’re trying to fix the Bazel build—which has nothing to do with it ...”} Such feedback may be technically justified, but it can halt progress if the author does not return with an alternative solution.
\\


\textbf{Conflict of Perspectives.} This category captures stalling caused by conflicting interpretations or priorities rather than by the absence of communication. It includes two related but distinct subcategories: \textit{Disagreement} (1.67\%) and \textit{Diverging Opinions} (0.93\%). Disagreement refers to cases where participants explicitly oppose or challenge one another’s position, such as whether to use a particular implementation strategy, whether a proposed API design is appropriate, whether a refactoring is worth the risk, or whether a change aligns with the project’s intended direction. Diverging Opinions, by contrast, captures cases where participants advance different preferences, priorities, or design directions without necessarily engaging in direct opposition. These cases are important because they cannot always be resolved through additional implementation effort alone. Instead, they often require persuasion, compromise, clarification of project goals, or a maintainer's decision to move the PR forward.

For example, in a QGIS PR\footnote{\url{https://github.com/qgis/qgis/pull/8023}}, the reviewer expressed conditional openness i.e., \textit{Diverging Opinions}, to the proposed change but withheld approval until the author provided a stronger rationale: \textit{“I’m not totally opposed to it if there is an ultimate benefit of it. I just need to have that one presented here first ;) ...”} This type of unresolved disagreement can leave a PR in an ambiguous state: it is not rejected, but it is also not approved.
\\


\textbf{Technical \& Testing Issues.} This category captures PRs that stall because the proposed change cannot yet be safely integrated due to unresolved technical problems, failing validation, or difficulty reproducing the reported behavior. It includes three subcategories: \textit{Technical Issues} (3.80\%), \textit{Unresolved Test Failures} (3.33\%), and \textit{Non-Reproducible Issues} (0.20\%). Technical Issues refer to implementation-level problems in the proposed change, such as incorrect logic, broken functionality, compatibility concerns, dependency conflicts, or design choices that require further correction before the PR can proceed. Unresolved Test Failures capture cases where automated checks, CI builds, or project tests fail, and the author or reviewers cannot immediately determine whether the failure is caused by the PR, the test suite, or the execution environment. Non-Reproducible Issues refer to cases where a reported bug, failure, or unexpected behavior cannot be reproduced by other participants, making it difficult to confirm the problem or evaluate the adequacy of a proposed fix. Unlike collaboration-related stalling, these cases often involve objective barriers to integration: the PR cannot move forward until the technical defect is fixed, the failing test is diagnosed, or the reported behavior is reproduced.

For example, in a Home Assistant PR\footnote{\url{https://github.com/home-assistant/core/pull/42344}}, the author reported that tests passed locally but failed after the PR was submitted: \textit{“... The tests pass locally using pytest ... but then fail when I submit the PR ...”} This example illustrates how unresolved test failures may arise not simply from incorrect code, but from discrepancies between local and CI environments. Such failures can keep a PR inactive even when the author remains engaged, because progress depends on diagnosing whether the problem lies in the submitted change, the test infrastructure, dependency versions, or CI-specific execution conditions.
\\

\textbf{Policy \& Procedural Issues.} This category captures non-technical blockers arising from project governance, compliance requirements, or administrative procedures. It includes \textit{Contributor License Agreement (CLA) Issues} (2.67\%), \textit{Administrative Issues} (1.13\%), and \textit{Redundant PRs} (0.07\%). These cases show that a PR may stall even when the proposed code is technically acceptable, because the contribution must satisfy project-specific rules before it can be merged.

CLA-related issues are particularly notable because their resolution often depends on actions outside the code review discussion itself. Redundant PRs represent a different procedural challenge: multiple contributors may submit overlapping solutions, requiring maintainers to decide which contribution should proceed. For example, in a Helm Charts PR\footnote{\url{https://github.com/helm/charts/pull/7670}}, the contributor acknowledged that another PR already addressed the same goal: \textit{“\#8094 accomplishes the same goal. Sorry for the duplicated work ...”} Such cases can leave one PR inactive while maintainers proceed with another.
\\

\textbf{External Constraints \& Limitations.} This category captures stalling caused by dependencies or conditions outside the project’s immediate authority. Although both this category and \textit{Policy and Procedural Issues} involve factors beyond implementation work, they differ in where the blocking authority resides. Policy and procedural issues are governed by the project’s own contribution rules, administrative requirements, and maintainer decisions, whereas external constraints depend on systems, actors, release processes, or related projects outside the project’s direct control. This category includes \textit{External Dependency} (3.80\%) and \textit{External Factors (0.67\%)}. These cases are distinct from internal technical or procedural blockers because the project cannot resolve them independently. Instead, progress depends on upstream libraries, external systems, third-party services, release schedules, infrastructure, or related projects.

For example, in a conda-forge PR\footnote{\url{https://github.com/conda-forge/staged-recipes/pull/10978}}, the author explained that the PR was waiting for another upstream PR to be released before it could proceed: \textit{“This PR is waiting on [upstream PR link] to be released. Once that happens, this PR can be updated for the new release ...”} This illustrates how dependency chains in modern software ecosystems can cause a PR to remain inactive even when both authors and maintainers are aware of the required next step. In such cases, the PR is not stalled because the contribution violates an internal policy or lacks a maintainer decision, but because the next enabling action must occur outside the project’s own review process.

\subsubsection{Inline Review Comment}\label{inline review}
To complement the PR-level perspective provided by general review comments, we also analyzed inline review comments. Inline review comments offer a more localized view of PR inactivity because they are attached to specific code fragments or diff hunks. As a result, they reveal concrete implementation-level blockers that may not be visible from PR-level general discussions. Our analysis of inline comments identified five high-level categories and 17 subcategories of stalling factors.
\\

\textbf{Review Process \& Communication Issues.} As with general review comments, communication-related issues are the largest category in inline comments. However, their nature is different. In general comments, communication problems often appear as broad unresponsiveness or lack of follow-up. In inline comments, they more often appear as unresolved code-level discussions, pending approval, or repeated review iterations.

The most frequently observed subcategory is \textit{Unresolved Discussion} (18.39\%), which captures cases where participants discuss a specific code change but do not reach a decision. This is different from simple unresponsiveness: the discussion may contain active engagement, but the PR stalls because the participants do not converge on what should be done. For example, in one inline thread\footnote{\url{https://tinyurl.com/mh83vp9z}}, the author expressed willingness to parameterize a behavior but left the final decision to the reviewer: \textit{“I am open to parameterizing it. Rolling update is the default, so it seemed unnecessary to keep. I can put it back if explicitness is desired.”} Without a definitive response, the discussion remains unresolved. In addition, \textit{Final Approval Pending} (14.39\%) captures cases where substantive issues appear to have been addressed, but the PR still awaits a formal approval or merge action. This finding suggests that inactivity can occur even near the end of the review process. \textit{Communication Delays} (8.75\%) and \textit{Review Process Challenges} (5.79\%) further show that inline review can stall because of delayed responses, repeated iterations, reviewer availability, large PRs, or shifting project priorities.
\\


\textbf{Technical \& Dependency Issues.} This category captures technical blockers surfaced through code-level discussions. It includes \textit{Dependency and Configuration Problems} (8.95\%), \textit{General Technical Issues} (5.72\%), \textit{Testing and Environment Failures} (4.06\%), \textit{Versioning and Compatibility Issues} (2.48\%), and \textit{Security and Compliance Issues} (1.24\%). Together, these subcategories show that inline discussions often stall when a specific code change raises unresolved concerns about dependencies, configuration, build behavior, test execution, runtime environments, compatibility, or security requirements. Compared with general review comments, inline comments expose these blockers with greater specificity because they are tied to concrete artifacts such as package files, configuration scripts, tests, imports, and implementation details.

For instance, in one PR\footnote{\url{https://tinyurl.com/3jykaj44}}, the author noted that an upstream dependency issue still needed to be reported: \textit{“... There’s still lua-zlib’s ridiculous behaviour to report upstream to lua-zlib as well ...”} This comment illustrates how an inline discussion can reveal both the immediate technical concern and the external dependency that prevents the PR from progressing.
\\


\textbf{Code Quality Issues.} Code quality issues appear prominently in inline comments because they are naturally tied to specific code locations and implementation details. This category includes \textit{Styling and Formatting Issues} (7.71\%), \textit{Code Structure and Optimization Problems} (4.96\%), and \textit{Documentation Gaps and Inconsistencies} (4.82\%). Unlike technical failures that directly break builds or tests, these issues often concern whether the submitted code is readable, maintainable, consistent, and understandable enough to be integrated into the project. Although such issues may not always affect immediate functionality, they can still delay merging when reviewers require additional revisions to improve naming, formatting, documentation, code organization, or alignment with project conventions.

The most frequent subcategory, \textit{Styling and Formatting Issues}, shows that even relatively small code-level concerns can contribute to PR inactivity. These issues include inconsistent formatting, naming inconsistencies, typos, and deviations from established style conventions. While individually minor, they often require author updates and subsequent reviewer confirmation. When such changes are not addressed promptly, or when they trigger additional rounds of review, the PR may remain inactive despite the absence of major functional defects. \textit{Code Structure and Optimization Problems} capture more substantive quality concerns related to how the code is organized, refactored, or optimized. These comments often point to opportunities to simplify logic, reduce duplication, improve readability, or restructure the implementation for maintainability. Compared with formatting issues, these concerns may require deeper changes because they can affect the design and long-term evolution of the codebase. As a result, the PR may stall while the author revises the implementation or while reviewers decide whether the current structure is acceptable. Similarly, \textit{Documentation Gaps and Inconsistencies} represent another important source of stalling. These comments involve missing explanations, incomplete documentation, inconsistent naming, or localization-related problems. Documentation concerns may seem secondary to functionality, but reviewers often treat them as necessary for future maintainability, especially when the code introduces non-obvious behavior, temporary workarounds, or dependencies on upstream issues. For example, in one inline comment\footnote{\url{https://tinyurl.com/5rdtpzd2}}, the reviewer requested an explanatory documentation and a link to an upstream issue: \textit{“... Could you also add a comment with a link to the upstream issue, so we know when to remove this line?”} Although the requested change is small, it requires author action and subsequent reviewer confirmation.
\\

\textbf{Implementation \& Decision-Making Challenges.} This category captures cases where progress depends on resolving implementation choices, design trade-offs, architectural disagreements, or incomplete work. It includes \textit{Design and Architecture Conflicts} (6.06\%), \textit{Implementation Challenges} (3.17\%), and \textit{Performance and Efficiency Issues} (0.69\%). These issues differ from routine code-quality concerns because they often require agreement on the direction of the solution, not merely a localized edit.

\textit{Design and Architecture Conflicts} are the most frequent subcategory in this group. These conflicts may require input from maintainers or broader project consensus, especially when the proposed change affects long-term maintainability or project structure. For instance, in one PR\footnote{\url{https://tinyurl.com/2km939kz}} involving installation strategy, the author summarized the unresolved decision: \textit{“We need to decide whether we stick with the current approach or restructure the installation process entirely. The build script is causing complications, and we need to finalize a solution.”} Such cases can stall because there is no straightforward code edit until the project agrees on the preferred approach.
\\


\textbf{Merging \& Review Delays.} This category captures late-stage blockers that prevent a PR from being finalized even after much of the substantive review discussion has already taken place. It includes \textit{Iterative Review and Refinements Delay} (1.86\%) and \textit{Conflicts and Merge Issues} (0.96\%). Although this is the least frequent high-level category in our inline-comment analysis, it is important because it shows that PR inactivity can occur not only when participants disagree about the main contribution, but also when the PR remains trapped in the final stages of review, revision, or integration.

\textit{Iterative Review and Refinements Delay} refers to cases where the PR continues to receive small but unresolved requests for changes. These comments often concern minor corrections, local refinements, or follow-up edits that emerge after earlier review rounds. Individually, such requests may appear easy to address. However, they can still delay completion because each refinement requires author action, a new commit, and subsequent reviewer confirmation. In this sense, the PR is not blocked by a major technical defect, but by an unfinished review loop. The accumulation of small pending refinements can therefore keep a PR inactive, particularly when reviewers return with additional comments after each update or when the author does not promptly complete the final requested changes. \textit{Conflicts and Merge Issues} capture integration-related blockers, including merge conflicts, redundant changes, or restrictions on which parts of the diff should be modified. These issues differ from general technical problems because they arise at the boundary between the proposed change and the evolving project codebase. A PR may be technically acceptable in isolation, but remain unmerged because the branch is out of date, the change overlaps with another contribution, or the author modified generated, or protected sections of the code. Resolving these cases often requires rebasing, removing unintended edits, restoring generated files, or coordinating with maintainers about what should be included in the final diff. For example, in one PR\footnote{\url{https://tinyurl.com/3mmhhe7f}}, a reviewer instructed the author not to edit a specific generated block: \textit{“Please don’t edit the bottle block.”} Although this feedback is concise, it introduces an additional integration constraint: the author must revise the PR to avoid modifying a section that should remain generated, protected, or maintained through a different process. The PR therefore cannot simply proceed to merge, even if the main implementation is otherwise acceptable. It must go through another revision and confirmation cycle before the review can be completed.

\subsubsection{General vs.\ Inline Review Comments}

Taken together, the above analyses (Section \ref{general review} and \ref{inline review}) show that general and inline review comments provide complementary explanations of PR inactivity. General comments primarily expose PR-level breakdowns in review coordination, whereas inline comments reveal localized blockers embedded within specific code changes. This distinction is consistent with the broader role of modern code review as both a coordination mechanism and a quality-assurance practice, where reviewers assess not only whether a contribution should be accepted, but also how the proposed changes should be understood, revised, and integrated into the project \cite{bacchelli2013expectations, gousios2014exploratory}. Consequently, the same broad outcome, PR inactivity, appears differently depending on whether the discussion occurs at the level of the whole PR or at the level of a particular diff hunk. The main commonality across the two analyses is that communication remains central to PR inactivity. However, the form of communication-related stalling differs across comment types. In general review comments, inactivity is more often associated with interrupted review momentum, where the next required action is not taken by the author, reviewer, or maintainer, reflecting disengagement or unclear ownership~\cite{li2021you}. In inline comments, inactivity more often reflects unresolved coordination around a specific code-level issue. In other words, general comments tend to reveal stalling through disengagement or unclear ownership, while inline comments often reveal stalling through discussion without closure. This distinction is important because prior work has shown that review latency and PR outcomes are shaped not only by technical characteristics, but also by the responsiveness and timing of review interactions \cite{zhang2022pull, khatoonabadi2023wasted}.

Moreover, the two comment types differ in the granularity of technical information they expose. General comments identify technical, procedural, and external blockers at a broader level, such as unresolved validation failures, project governance requirements, or upstream dependencies. Inline comments make these blockers more concrete by tying them to specific artifacts, such as configuration files, test cases, dependency declarations, documentation, or implementation details. This aligns with prior work showing that code review contributes to software quality through localized inspection of code changes, review coverage, and maintainability-related feedback \cite{mcintosh2016empirical, rigby2013convergent}. Therefore, relying only on general review comments would provide an incomplete view of PR inactivity, particularly for cases where the blocking issue is embedded in the code rather than visible in the overall PR conversation. Another distinction concerns the type of action required to recover a stalled PR. General comment stalling often requires process-oriented intervention, such as assigning a reviewer, prompting an inactive participant, clarifying ownership, resolving administrative requirements, or waiting for an external dependency. Inline comment stalling often requires code or decision-oriented intervention, such as revising an implementation, resolving a configuration issue, addressing a maintainability concern, settling a design disagreement, or completing a final approval loop. This suggests, similar to prior studies~\cite{khatoonabadi2023wasted, li2021you}, that stale PRs should not be treated as a single homogeneous class of abandoned contributions. Some PRs require renewed human coordination, while others require technical correction, architectural decision-making, or explicit maintainer approval.

These findings have implications for both tooling and project governance. For PR-level inactivity, projects may benefit from mechanisms that make responsibility and next actions visible, such as reviewer reassignment, inactivity reminders, ownership tracking, and escalation policies. Prior work on stale bots and review bots suggests that automated support can help projects manage unresolved PRs \cite{khatoonabadi2023understanding, wessel2021don}. However, such mechanisms must be used carefully because closing or nudging inactive PRs without understanding the underlying cause may also negatively affect contributor experience and long-term community participation~\cite{wessel2022quality, steinmacher2015social}. For inline comments inactivity, tools should provide better visibility into unresolved discussions, pending design decisions, unaddressed code-quality concerns, dependency blockers, and final approval states. Rather than only detecting that a PR has become inactive, review tools should help diagnose whether the PR is waiting for a person, a technical fix, an external dependency, a design decision, or a final merge action.

\begin{mdframed}[
    linecolor=black!60,
    linewidth=1.5pt,
    backgroundcolor=yellow!8,
    innertopmargin=6pt,
    innerbottommargin=6pt
]

\noindent\textbf{\underline{RQ2 Findings:}} Our qualitative analysis identified \textbf{22 subcategories} across six categories from general review comments and \textbf{17 subcategories} across five categories from inline review comments. Across both comment types, \textbf{communication failures} are the dominant driver of PR inactivity, accounting for \textbf{76.60\%} of general comment stalling (e.g., unresponsiveness, lack of interaction, missing follow-up) and \textbf{47.31\%} of inline comment stalling (e.g., unresolved discussions, final approval delays). \textbf{Technical and dependency issues} (22.45\% in inline comments) and \textbf{code quality concerns} (17.49\%, including formatting and documentation gaps) emerge as significant secondary causes, surfacing with greater specificity in inline comments than in general comments. Notably, general comments reveal stalling through \textit{disengagement and unclear ownership}, while inline comments reveal stalling through \textit{discussion without closure}, suggesting that inactive PRs represent a heterogeneous class of problems requiring different recovery strategies: some demand renewed human coordination, others require technical correction, design decisions, or explicit maintainer approval.
\end{mdframed}

\subsection{RQ$_3$: Attribution and Consequences of PR Inactivity} \label{causeImpact}

\subsubsection{Attribution of PR Inactivity} 

To examine who is most often associated with PR inactivity, we mapped each stalling rationale identified in RQ$_2$ to the actor or condition most directly linked to the next required action. We distinguish four attribution groups: author-related, reviewer-related, shared author-reviewer responsibility, and system or external factors. This attribution should not be interpreted as blame assignment. Rather, it identifies where the immediate blockage appears to reside based on the review discussion. For example, an author-unresponsive case is attributed to the author because progress depends on the author returning to the PR, whereas a final-approval-pending case is attributed to the reviewer or maintainer because progress depends on a review or merge decision. This framing is consistent with prior work showing that PR outcomes and latency are shaped by both contribution characteristics and the dynamics of review interactions~\cite{zhang2022pull, khatoonabadi2023wasted}.

The results show that PR inactivity is rarely attributable to one side alone. In both general and inline review comments, shared author-reviewer responsibility accounts for the largest portion of stalling rationales, representing 35.07\% of general-comment cases and 36.02\% of inline-comment cases. These shared cases include situations such as no interaction, communication delays, unresolved discussions, and review-process challenges, where progress depends on reciprocal action rather than a single participant. This finding suggests that many stalled PRs are better understood as coordination failures within the review process, rather than as failures of only authors or only reviewers. At the same time, the distribution of responsibility differs across comment types. In general review comments, reviewer-related factors account for 32.86\% of stalling rationales, compared with 20.27\% for author-related factors. This pattern suggests that, at the PR level, many stalled contributions are waiting for reviewer or maintainer action, such as providing feedback, responding to updates, redirecting the PR, requesting changes, or making a final review decision. This does not mean that reviewers are solely responsible for PR inactivity. Rather, it indicates that the broader review process often depends on reviewer-side actions to keep the PR moving. In inline review comments, the attribution pattern is more balanced. Author-related factors account for 21.35\% of cases, while reviewer-related factors account for 20.18\%. This balance reflects the nature of inline review, where progress often depends on both the author addressing code-level feedback and the reviewer confirming whether the revision is acceptable. For example, code-quality issues, implementation challenges, and unresolved technical concerns often require author action, whereas unresolved discussions and pending approval require reviewer or maintainer response. Inline comments therefore reveal a more reciprocal form of responsibility, where inactivity can emerge from either delayed implementation work or delayed review closure. System-related and external factors also play an important role, especially in inline comments, where they account for 22.45\% of cases. These factors include testing failures, dependency and configuration problems, compatibility issues, security concerns, and other technical blockers that cannot always be attributed directly to either the author or reviewer. Their presence shows that PR inactivity is not only a human coordination problem. In many cases, progress is constrained by CI failures, dependency behavior, environment inconsistencies, or upstream changes. This finding reinforces the socio-technical nature of PR inactivity: even when participants remain engaged, the PR may remain stalled because the surrounding technical infrastructure prevents closure.

Overall, the attribution analysis reveals that stalled PRs are most often a shared responsibility, but the locus of responsibility changes depending on the level of review. General comments highlight reviewer and maintainer roles in sustaining review momentum, while inline comments reveal a more balanced interaction between author-side revisions, reviewer-side closure, and system-level blockers. Therefore, interventions should avoid assigning blame to one actor. Effective interventions should therefore avoid assigning blame to a single actor. Instead, review tools should identify the \textit{next required action} and surface it to the appropriate participant, whether
that action involves author revision, reviewer response, maintainer approval, or resolution of a technical dependency.



\begin{table*}[htbp]
\centering
\caption{Impact on Collaboration due to PR Stalling.}
\label{table 5}
\resizebox{11cm}{!}{%
\begin{tabular}{lrr}
\toprule
\multicolumn{1}{l}{\textbf{Project}} & \multicolumn{1}{c}{\textbf{Author Retention (\%)}} & \multicolumn{1}{c}{\textbf{Re-Engagement (\%)}} \\ \toprule
apache-beam                          & 48.73                                              & 21.14                                           \\
wp-calypso                           & 57.86                                              & 20.00                                           \\
ceph-ceph                            & 51.52                                              & 36.15                                           \\
cataclysm-dda                        & 50.67                                              & 9.63                                            \\
staged-recipes                       & 31.41                                              & 17.03                                           \\
devextreme                           & 44.12                                              & 50.00                                           \\
frappe-erpnext                       & 67.55                                              & 36.31                                           \\
grafana-grafana                      & 38.58                                              & 37.15                                           \\
grpc-grpc                            & \textbf{21.31}                                              & 28.57                                           \\
helm-charts                          & 24.76                                              & \textbf{4.67}                                            \\
home-assistant-core                  & 39.61                                              & 36.15                                           \\
home-assistant.io                    & 49.38                                              & \textbf{54.20}                                           \\
homebrew-core                        & 41.43                                              & 21.07                                           \\
istio-ishtio                         & 52.52                                              & 21.47                                           \\
nixos-nixpkgs                        & 48.29                                              & 10.96                                           \\
qgis-qgis                            & 60.61                                              & 39.29                                           \\
riot-os-riot                         & 33.93                                              & 26.17                                           \\
solana-labs-solana                   & 63.51                                              & 50.75                                           \\
wikia-app                            & \textbf{78.72}                                              & 32.26                                           \\ \midrule
\textbf{Total (\%)}                  & \textbf{39.56}                                     & \textbf{20.99}                                  \\ \bottomrule
\end{tabular}}
\end{table*}

\subsubsection{Impact on the Contribution Process} 


We next examined whether PR stalling is associated with reduced future participation and weaker reviewer-author continuity. Among the 7,042 contributors who experienced a stalled PR, only 39.56\% returned to make further contributions, meaning that 4,256 contributors did not contribute again within our observation window. This low retention rate suggests that stalled PRs may be associated with contributor disengagement, particularly when contributors do not receive timely feedback, clear next steps, or closure on their submitted work. Prior work on abandoned PRs similarly shows that unresolved or prolonged review processes can waste contributor and maintainer effort and may affect continued participation~\cite{khatoonabadi2023wasted}.

The project-level results in Table~\ref{table 5} show considerable variation in author retention. Some projects retain a relatively large proportion of contributors after a stalled PR, such as \textit{wikia-app} with 78.72\%, \textit{frappe-erpnext} with 67.55\%, \textit{solana-labs-solana} with 63.51\%, and \textit{qgis-qgis} with 60.61\%. In contrast, other projects show much lower retention, such as \textit{grpc-grpc} with 21.31\%, \textit{helm-charts} with 24.76\%, \textit{staged-recipes} with 31.41\%, and \textit{riot-os-riot} with 33.93\%. This variation suggests that the consequences of PR stalling may depend on project-specific factors, such as community responsiveness, contributor onboarding practices, review norms, project size, maintainer availability, or the extent to which stalled contributions receive clear explanations. These interpretations are consistent with prior work on barriers faced by open-source contributors and the role of social and process factors in sustaining participation~\cite{steinmacher2015social}.

Reviewer-author re-engagement shows an even more limited pattern of continuity. Across the 19 projects, only 20.99\% of reviewers re-engaged with the same contributor in the contributor's subsequent ten PRs. This indicates that, after a PR stalls, the same reviewer-author pair rarely continues to interact in later contributions. In some projects, re-engagement is particularly low, such as \textit{helm-charts} with 4.67\%, \textit{cataclysm-dda} with 9.63\%, and \textit{nixos-nixpkgs} with 10.96\%. In contrast, projects such as \textit{home-assistant.io}, \textit{solana-labs-solana}, and \textit{devextreme} show substantially higher re-engagement rates, exceeding or approaching 50\%. These results suggest that PR stalling may affect not only whether contributors return, but also whether the same reviewer-author relationship continues. Low reviewer re-engagement may reflect several possible mechanisms. Reviewers may not be assigned to the same contributor again, contributors may shift to different parts of the project, or stalled PRs may weaken the continuity of collaboration between the same participants. Regardless of the exact mechanism, the low overall re-engagement rate indicates that stalled PRs are often followed by limited continuity in reviewer-author interaction.

Importantly, these results should be interpreted as associations rather than causal evidence. We do not claim that PR stalling alone causes contributors to leave or prevents reviewers from re-engaging. Contributor retention and reviewer assignment can be influenced by many factors, including project governance, contributor motivation, task availability, reviewer workload, and the nature of the original contribution \cite{steinmacher2015social, soares2018factors, zhang2022pull}. However, the observed retention and re-engagement patterns indicate that stalled PRs are associated with weaker contribution continuity. This makes inactivity a concern not only for individual PR throughput, but also for long-term community sustainability.

Together, the attribution and impact analyses reveal that PR inactivity carries both process-level and community-level consequences. At the process level, stalled PRs reflect unresolved ownership of the next required action. At the community level, they are associated with reduced contributor return and limited reviewer-author continuity. These findings highlight the need for project practices that make stalled PRs actionable before they become permanently abandoned: assigning clear ownership, identifying whether the next action belongs to the author, reviewer, maintainer, or system, and providing timely explanations when progress is blocked. Prior work on stale bots suggests that automated support can help projects manage inactive PRs; however, such tools should be designed carefully, as automated closure or nudging without understanding the underlying cause may negatively affect contributor engagement and community participation~\cite{khatoonabadi2023understanding, wessel2021don}.


\begin{mdframed}[
    linecolor=black!60,
    linewidth=1.5pt,
    backgroundcolor=yellow!8,
    innertopmargin=6pt,
    innerbottommargin=6pt
]
\noindent\textbf{\underline{RQ$_3$ Findings:}} PR inactivity is not simply an author-side or reviewer-side problem, but often reflects unresolved ownership of the next action. Shared author-reviewer responsibility is the most common attribution pattern in both general and inline comments, while PR-level discussions more often depend on reviewer or maintainer action and inline discussions involve a more distributed mix of author revisions, reviewer closure, and system-level blockers. Stalled PRs are also associated with weaker contribution continuity, with only \textbf{39.56\%} of contributors returning after a stalled PR and only \textbf{20.99\%} reviewer-author re-engagement across subsequent submissions. These findings suggest that reducing PR inactivity requires clarifying responsibility, preserving review continuity, and supporting timely intervention before stalled PRs become abandoned.
\end{mdframed}

\section{Discussion}

\subsection{Contribution Type as a Proxy for Socio-technical Complexity}

The finding that feature enhancement and upgrading (54.04\%) and issue fixing (23.63\%) together account for over 77\% of stalled PRs (Table \ref{tab:category-distribution}) cannot be assumed to follow directly from their overall prevalence in the project ecosystem. A more informative interpretation is that many open source projects apply a largely uniform intake and review process across contribution types, even though the coordination infrastructure required to shepherd a complex, cross-cutting feature through review differs substantially from what is needed for a minor documentation update. Feature enhancements frequently require reviewers to assess design trade-offs, long-term maintenance implications, and alignment with project direction. These activities impose socio-technical coordination costs that typical review tooling only partially supports, making such PRs more vulnerable to delays and stalled decision-making \cite{yang2026roadmap, yu2015wait}.

This interpretation aligns with and extends prior findings on PR evaluation and abandonment. Studies of abandoned PRs report that reviewer unresponsiveness, difficulty reaching consensus, and obstacles introduced during the review process are recurring reasons for abandonment, particularly when contributions require prolonged iteration and negotiation \cite{li2021you, khatoonabadi2023understanding}. Yet these studies generally operationalise complexity through aggregate indicators or retrospective explanations rather than linking stalling risk to contribution type at fine granularity. By providing type-level resolution, our classification results suggest that stalling concentrates in contribution categories that demand sustained reviewer engagement and decision-making. A practical implication is that a single uniform queue can structurally delay high coordination work by forcing feature and issue PRs to compete for attention with lower coordination changes, even though their completion depends on earlier and more consistent reviewer commitment. This finding also resonates with the broader concern regarding reviewer assignment. Work on PR latency shows that process-related factors such as reviewer workload and integrator responsiveness materially influence evaluation timelines \cite{yu2015wait}. When complex feature PRs enter a project without an explicitly committed reviewer and without type-aware triage, they are placed in direct competition with simpler PRs for the same constrained reviewer pool. This queue design can systematically disadvantage contributions that are most vulnerable to stalling and can amplify the risk that high-value changes drift into inactivity.

\subsection{Communication Breakdown as a Dominant Stalling Pathway}
The dominant stalling mechanism across both general and inline review comments is communication breakdown, including no interaction, author unresponsive, reviewer unresponsive, and missing feedback (Tables \ref{table 2} and \ref{table 4}). A narrow reading would treat these as incidental coordination failures. A more informative interpretation is that silence functions as a socially meaningful signal within pull-based collaboration. Under prolonged ambiguity, unanswered questions, or uncertain acceptance criteria, participants may reduce effort or disengage when the perceived likelihood of progress declines. This interpretation is consistent with empirical studies of PR abandonment that identify delayed responses, difficulty reaching agreement, and review process friction as recurrent pathways to stalled progress and eventual abandonment \cite{li2021you, khatoonabadi2023wasted}. It also aligns with motivation theories that emphasise the role of timely feedback and perceived competence and relatedness in sustaining voluntary effort \cite{deci2000and}.

Separating causes at the general and inline comment levels provides additional insights. Communication breakdown and technical blockage appear as parallel stalling pathways rather than a single sequential chain. Inline-level technical dependency issues account for a substantial share of stalling (Table \ref{table 4}), including failed checks, dependency mismatches, and environment issues. This suggests that some stalled conversations may reflect low agency technical states in which contributors cannot easily diagnose infrastructure failures without maintainer guidance, rather than interpersonal disengagement. Prior work has documented that CI and integration constraints can materially shape PR outcomes and review progress, and that abandonment can follow from obstacles contributors face during the review process \cite{yu2015wait, khatoonabadi2023wasted}. The practical consequence is interpretive: inactivity metrics based only on comment latency can conflate technically blocked PRs with socially disengaged ones, even though they represent different underlying states.

From the satisfaction perspective, both pathways converge on the same experiential outcome. Contributors receive little confirmation that progress is being made, limited actionable direction, and weak social reinforcement that the effort is valued. In self-determination terms, prolonged absence of feedback can erode perceived competence and relatedness, which are central to sustained motivation in collaborative settings \cite{deci2000and}. In other words, stalled PRs represent not only delayed work but also a risk to contributor retention, as silence and blockage reduce the perceived payoff of continued participation.

\subsection{Reviewer Responsiveness as a Structural Coordination Constraint}
The attribution analysis (Section \ref{causeImpact}) indicates that reviewer-side responsibility for stalling is larger than author-side responsibility in general review comments (32.86\% versus 20.27\%), while shared responsibility constitutes the largest category (35.07\%). These distributions suggest that stalling cannot be interpreted solely as contributor withdrawal. Instead, it reflects a coordination process in which control over pacing and decision making is structurally asymmetric. In pull-based development, reviewers and maintainers gate progress through response timing, change request scope, and merge decisions, while contributors often enter a waiting state after investing implementation effort. Prior research on PR abandonment similarly reports that delayed responses and difficulty reaching agreement are common antecedents of abandonment, and that abandonment frequently follows obstacles encountered during review rather than purely contributor-side factors \cite{li2021you, khatoonabadi2023wasted}.

This asymmetry is amplified in open-source contexts because expectations of responsiveness are often governed by informal norms rather than by explicit service targets or enforced policies. When review progress depends on a small number of overloaded maintainers, delays can emerge even in the absence of negative intent. Evidence on PR evaluation latency shows that process and workload conditions materially contribute to slow reviews and prolonged waiting times \cite{yu2015wait}. From the contributor's perspective, a prolonged absence of reviewer engagement can be interpreted as uncertainty about whether the contribution remains viable, increasing the opportunity cost of continued effort and making disengagement more likely \cite{li2021you}. In this sense, stalling highlights not only communication quality issues but also the structural constraints of maintainer capacity and attention.

Moreover, the low reviewer re-engagement rate after a stalling event (Table \ref{table 5}) further suggests that stalled PRs can disrupt future collaboration trajectories. When reviewers do not return to the same contributor after a stalling episode, the relationship may not recover even if the immediate PR is closed. Prior work characterizes such outcomes as wasted contributions, where effort is expended but fails to translate into integration and sustained participation \cite{khatoonabadi2023wasted}. Therefore, the central implication is interpretive: reviewer responsiveness should be understood as a community-level coordination property rather than only an individual behavior, because it shapes whether contributors perceive continued participation as likely to pay off.

\subsection{Stalling as an Outcome of Accumulated Coordination Debt}
Across categories and comment levels, this study shows that stalling is rarely caused by a single isolated factor. Instead, stalling emerges when the demand for coordination accumulates faster than reviewers' attention and decision-making capacity, producing prolonged uncertainty about next actions and eventual silence. This interpretation aligns with research on PR abandonment that highlights delayed responses, difficulty reaching consensus, and obstacles introduced during review as recurring pathways to abandonment \cite{li2021you, khatoonabadi2023wasted}.

The central contribution of this study is to connect stalling to developer satisfaction and community sustainability. When contributors cannot obtain timely confirmation, actionable guidance, or clear closure, the perceived payoff of continued participation declines, increasing the likelihood of non-return \cite{li2021you}. This mechanism-level framing explains why uniform queueing and threshold-only automation can be insufficient under heterogeneous blocking states, and why projects with constrained maintainer capacity are especially vulnerable to silent attrition~\cite{yu2015wait}.

\subsection{Operational Guidelines for Minimizing PR Stalling}

Our findings show that PR stalling is not caused by a single failure mode. Instead, stalled PRs reflect a heterogeneous set of socio-technical barriers, including communication breakdowns, unresolved technical blockers, unclear ownership of the next action, and limited collaboration continuity. Across general review comments, collaboration and availability issues dominate, while inline comments reveal more localized blockers such as unresolved discussions, dependency and configuration problems, testing failures, and code-quality concerns. In addition, our attribution analysis shows that stalled PRs are often a shared author-reviewer problem rather than the responsibility of a single actor. These findings suggest that projects need both general contribution-management practices and more targeted mechanisms that diagnose why a PR is inactive before applying an intervention.
\\

\subsubsection{Clarify Contribution Expectations and Review Ownership} 
Because many stalled PRs in our dataset were associated with no interaction, missing feedback, no follow-up, and unclear ownership of the next required action, projects should make contribution expectations explicit from the beginning. A detailed and well-structured \textit{CONTRIBUTING.md} file can specify acceptable contribution types, expected response windows, review stages, escalation paths, and what contributors should do when a review becomes inactive. Such guidance is particularly important because our results show that feature enhancement and issue-fixing PRs constitute the majority of stalled cases, and these PRs often require sustained discussion about design choices, implementation direction, and project priorities. Prior work similarly shows that inconsistencies between documented guidelines and actual project practices can discourage participation and increase abandonment \cite{elazhary2019not}, while clearly defining project scope and acceptable contribution types can help align contributor expectations with maintainer priorities \cite{coelho2017modern, kobayakawa2017github}. Thus, contribution documentation should not only describe how to submit a PR, but also clarify who is expected to act next when a PR is waiting for author revision, reviewer feedback, maintainer approval, or external resolution.
\\


\subsubsection{Adopt Type-aware and Proactive PR Triage} 
Our PR-type analysis indicates that feature enhancement and upgrading PRs, followed by issue-fixing PRs, represent the largest portion of stalled contributions. These PRs may be more coordination-intensive than simple documentation or formatting changes because they often require design evaluation, consensus building, testing, and long-term maintainability assessment. Therefore, projects should avoid treating all PRs as equivalent items in a single review queue. Instead, maintainers can use type-aware triage to identify PRs that are likely to require deeper coordination, assign an appropriate reviewer early, and clarify whether the proposed change aligns with project priorities before contributors invest substantial additional effort. General contribution-management practices such as issue and PR templates can support this process by requiring contributors to provide motivation, expected behavior, test evidence, dependency changes, and design trade-offs at submission time \cite{zhang2022consistent, sulun2024empirical}. Promptly addressing non-compliant or out-of-scope PRs with clear and respectful feedback can also reduce wasted effort while preserving contributor motivation \cite{ferreira2021shut, rahman2024words}. For newcomers, labels such as ``good first issue'' and mentorship structures can further improve onboarding and reduce avoidable abandonment \cite{steinmacher2021being, horiguchi2021onboarding}.
\\


\subsubsection{Make the Next Required Action Explicit}
A central implication of our attribution analysis is that stalled PRs often persist because the next action is unclear or distributed across multiple participants. Shared author-reviewer responsibility is the largest attribution pattern in both general and inline comments, while reviewer-side responsibility is particularly visible in PR-level discussions. Therefore, review systems and project practices should explicitly identify whether a PR is waiting for author revision, reviewer response, maintainer approval, CI repair, dependency release, design decision, or final merge action. This is different from simply marking a PR as inactive. For example, an author-unresponsive PR requires a different intervention from a PR that is waiting for final approval, blocked by CI failures, or dependent on an upstream release. Projects can operationalize this through lightweight status labels such as \textit{awaiting author}, \textit{awaiting reviewer}, \textit{awaiting maintainer decision}, \textit{blocked by CI}, \textit{blocked by dependency}, or \textit{needs design decision}. Such labels can reduce ambiguity, help contributors understand what is expected, and help maintainers prioritize stalled PRs more effectively.
\\

\subsubsection{Improve Review Feedback and Bottleneck Handling} 
Our general comment analysis shows that missing feedback, no follow-up, reviewer unresponsiveness, and change requests contribute to PR stalling. Inline comments further show that unresolved discussions and final approval delays are common sources of inactivity. These findings suggest that reviewers should provide feedback that is not only technically correct, but also actionable and closure-oriented. Incomplete or non-optimal PRs should be met with clear, constructive, and unambiguous feedback that explains the rationale for the requested change, identifies the specific next step, and provides documentation links or examples when possible. Prior studies suggest that such feedback can improve contributor retention, PR quality, and reduce repeated invalid submissions \cite{steinmacher2018almost, gottigundala2021qualitatively, papadakis2020did, golzadeh2019effect}. To reduce reviewer-side bottlenecks, projects can also distribute review responsibilities through team-based assignments, backup reviewers, and domain-specific reviewer rotation, especially for complex feature and issue-fixing PRs \cite{al2020workload, hajari2024factoring}. This recommendation is directly connected to our finding that stalled PRs are often waiting for reviewer or maintainer action, particularly in general review discussions.
\\


\subsubsection{Separate Technical Blockage from Social Disengagement}
Our inline comment analysis shows that PR inactivity is not always a sign of contributor disengagement. Many stalled PRs involve technical and dependency issues, including testing and environment failures, dependency and configuration problems, versioning concerns, security and compliance issues, and general technical blockers. Therefore, projects should avoid treating every inactive PR as an abandoned contribution. A PR blocked by CI failure, dependency release, or environment inconsistency requires diagnostic and technical support, not merely a reminder to the author. Projects can reduce such technical stalling by maintaining reliable CI pipelines, documenting local test environments, requiring minimal reproducible evidence for bug fixes, and using automated checks to distinguish code-quality issues from infrastructure failures. This distinction is important because inactivity metrics based only on elapsed time can conflate socially disengaged PRs with technically blocked PRs, even though the recovery strategies differ substantially.
\\

\subsubsection{Optimize PR Scope and Early Quality Gates}
Our findings show that code-quality concerns, including styling and formatting issues, code-structure problems, optimization concerns, and documentation gaps, contribute substantially to inline-level stalling. These issues may not always be conceptually difficult, but they can trigger repeated review cycles and delay closure. Therefore, projects should encourage smaller, focused PRs and apply early quality gates before human review. Prior research suggests that smaller PRs are easier to review and can accelerate merge decisions \cite{ram2018makes, sadowski2018modern}. Automated linting, formatting checks, documentation checks, and test requirements can also filter out routine issues before reviewers spend attention on them \cite{singh2017evaluating, dos2018investigating, panichella2020empirical}. This allows reviewers to focus on substantive concerns such as correctness, design, maintainability, and architectural fit, while reducing avoidable review iterations caused by formatting, naming, or documentation inconsistencies \cite{wessel2022quality}.
\\


\subsubsection{Design Cause-aware Automation Rather than Time-only Stale Interventions}
Our results suggest that time-based stale detection is insufficient because stalled PRs have different underlying causes. Some PRs are waiting for author response, some are waiting for reviewer closure, some are blocked by technical infrastructure, and others require project-level decisions or external dependency updates. Therefore, automation should move beyond generic stale warnings and support cause-aware intervention. For example, an engagement bot such as Nudge \cite{maddila2023nudge} could issue different reminders depending on whether the discussion history indicates missing author action, pending reviewer approval, unresolved CI failure, or an upstream dependency blocker. Similarly, maintainer dashboards could categorize inactive PRs by stalling reason, making it easier to distinguish ``awaiting author response'' from ``CI failure,'' ``needs maintainer decision,'' or ``blocked by external dependency'' \cite{guizani2022attracting}. Backup reviewer assignment and reviewer rotation can also be triggered when a PR waits too long for review closure. However, automation should be used carefully. Since prior work shows that stale bots can create friction and reduce contributor engagement when applied indiscriminately, automated closure or nudging should be accompanied by explanation, context, and a clear path for reactivation. In this way, automation can support review coordination without turning inactivity management into a purely mechanical closure process.


\section{Research Implications}

\subsection{For Practitioners}

\subsubsection{Differentiate intake protocols by contribution complexity}
Project leads should avoid treating all PRs as equivalent units in a single uniform queue. In our dataset, feature-enhancement and upgrading PRs and issue-fixing PRs together constitute over 77\% of stalled cases (Table \ref{tab:category-distribution}). Although this distribution should not be interpreted as a stall likelihood without comparing against all PRs, it indicates that a large share of stalled work occurs in contribution types that often require design discussion, technical validation, and sustained reviewer engagement. For mature projects, a practical option is to introduce contribution triage lanes: a fast-track lane for low-risk changes such as documentation updates, dependency bumps, and minor fixes, and a managed-complexity lane for feature and issue-fixing PRs. In the managed-complexity lane, a reviewer can be explicitly assigned at submission time, expected response windows can be stated up front, and early maintainability or scope concerns can be clarified before contributors invest further effort. This design helps reduce the mismatch between documented contribution processes and actual review practices, a gap that prior work shows can discourage participation \cite{elazhary2019not}. It also responds to evidence that PR abandonment often follows obstacles introduced during review, including delayed feedback, unclear expectations, and difficult revision processes \cite{khatoonabadi2023wasted}.

\subsubsection{Treat reviewer responsiveness as a community health metric}
Our attribution and post-stale engagement analyses suggest that PR stalling is not only an author-side disengagement problem. Reviewer-side and shared author-reviewer factors account for a substantial portion of stalled cases, and reviewer re-engagement with the same contributor averages only about 21\% after a stalling event (Table \ref{table 5}). Projects should therefore monitor reviewer responsiveness as a community health signal rather than treating silence as a benign absence. Practical metrics include time-to-first-human-response, time-to-next-review-action, discussion-resolution rate, number of PRs waiting for reviewer action, and post-stale contributor return rate. These measures can complement conventional project metrics such as merge rate, issue-closure rate, and backlog size. Prior work on PR abandonment similarly highlights delayed responses and review-process friction as contributors to abandonment, which reinforces the need to surface these signals early enough for maintainers to intervene \cite{li2021you}.

\subsubsection{Redesign stale bot interventions as structured re-engagement prompts}
Many stale-bot deployments label and close inactive PRs using uniform inactivity thresholds, without identifying why progress stopped. Our findings show that this is problematic because stalled PRs reflect heterogeneous blocking states, including communication gaps, unresolved technical failures, unclear ownership of the next action, external dependencies, and unresolved design decisions. Therefore, stale automation should move from time-only closure toward reason-aware re-engagement. Instead of issuing the same stale message to every inactive PR, bots should classify the dominant blocking state and contact the actor most able to move the PR forward. For example, the bot could notify an author when checks are failing, request a reviewer update when feedback is pending, ask a maintainer to make a scope decision, or mark a PR as externally blocked when it depends on an upstream release. Prior work shows that stale automation can help reduce backlog, but can also create friction and reduce engagement when applied indiscriminately \cite{khatoonabadi2023understanding}. A reason-aware workflow can preserve the backlog-management benefits of automation while reducing the risk that still-viable contributions are silently closed.

\subsubsection{Apply stricter review comment clarity standards in coordination-intensive PRs}
Our inline-comment analysis shows that review-process and communication issues account for 47.31\% of inline stalling causes. This suggests that unclear direction, unresolved discussion, pending approval, and repeated review iterations are not merely short-term inconveniences, but can contribute to stalled PR conversations. Reviewers handling feature-enhancement, issue-fixing, or design-heavy PRs should therefore apply stronger clarity standards when requesting changes. A useful review comment should state what change is requested, why it is needed, how the author can validate completion, and whether the comment is blocking or optional. This recommendation aligns with prior evidence that useful review feedback tends to be specific, informative, and contextualized \cite{rahman2017predicting, turzo2024makes, rahman2025investigating}. It also reflects practitioners' preference for comments that explain what something does, how it should be used, and why it exists \cite{hu2022practitioners}.

\subsubsection{Prioritize CI-related and dependency blockers in stalled PR triage}
Our inline taxonomy shows that technical and dependency issues account for 22.45\% of inline stalling causes, including testing and environment failures, dependency and configuration problems, versioning and compatibility issues, security and compliance concerns, and general technical issues. These cases indicate that inactivity is not always caused by lack of motivation or lack of response. Some contributors may be unable to proceed because checks fail in CI, dependencies are unavailable, failures are flaky, or project-level infrastructure differs from the local environment. Projects should therefore adopt explicit CI triage protocols for stalled PRs. These protocols may include confirming whether a failure is deterministic or flaky, identifying whether the failure is caused by the PR or by unrelated infrastructure, providing reproduction steps, and assigning maintainer-side responsibility when the contributor lacks permission or context to resolve the issue. This reduces time spent in low-agency states where contributors cannot make progress despite being willing to continue \cite{khatoonabadi2023wasted, li2021you}.

\subsubsection{Make decision and acceptance criteria explicit to prevent indefinite negotiation}
Some stalled PRs are not blocked by implementation work, but by unresolved decisions, shifting requirements, or ambiguous acceptance thresholds. This is visible in our taxonomy through unresolved discussions, design and architecture conflicts, final approval pending, and iterative review delays. Projects should counter this by explicitly stating acceptance criteria when requesting changes and by recording a short decision outcome when discussions become design-focused. For example, maintainers can mark a PR as accepted with constraints, revision required with a concrete checklist, deferred with a rationale, or closed with a clear explanation. This practice transforms open-ended discussion into a bounded plan and reduces the likelihood that uncertainty evolves into silence. It also aligns with prior work showing that review-process friction and delayed convergence can contribute to abandonment \cite{khatoonabadi2023wasted, li2021you, yu2015wait}.

\subsection{For Researchers}

\subsubsection{Standardize stalling and abandonment operationalizations and benchmarking protocols}
Across PR abandonment and bot studies, ``stalling,'' ``abandonment,'' and ``stale closure'' are often operationalized using different inactivity thresholds and outcome rules, which makes findings difficult to compare and limits the transferability of predictive models. Future work should define benchmark-ready and intervention-aligned labels that separate inactivity from administrative closure. Researchers should also report sensitivity analyses across key inactivity thresholds, such as 30, 60, 90, and 180 days, and distinguish bot-triggered closure from human-initiated closure. Such benchmarks would enable fairer evaluation of risk models and interventions across projects with different review cadences, governance structures, and automation policies \cite{li2021you, khatoonabadi2023wasted, hasan2023understanding}.

\subsubsection{Build dual-channel, type-aware PR stalling risk models}
Future abandonment and stalling prediction should move beyond binary abandoned versus non-abandoned classifiers trained only on aggregate PR features. Our taxonomy motivates multi-class, dual-channel models that distinguish communication-driven stalling from technically driven stalling. Communication-driven stalling may be observable from response latency, discussion structure, missing feedback, unresolved comments, and linguistic signals in general review discussions. Technically driven stalling may be observable from CI outcomes, failing checks, dependency changes, configuration problems, and inline discussion topics. Separating these channels matters because their intervention requirements differ. Communication breakdowns call for workflow or social interventions such as targeted nudges, reviewer reassignment, or clearer ownership of the next action, whereas technical blockages call for engineering escalation such as CI triage, reproducibility guidance, or infrastructure fixes. Existing research and platform support typically operationalize stalling using coarse inactivity thresholds or aggregate predictors, and rarely expose reason-specific classifications that map directly to intervention design \cite{khatoonabadi2023understanding, khatoonabadi2023wasted}.

\subsubsection{Conduct longitudinal studies of contributor re-entry and community recovery}
Our reviewer's re-engagement rate of about 21\% after a stalling event provides an initial signal of weakened collaboration continuity, but it cannot distinguish durable relationship degradation from temporary maintainer capacity constraints. Longitudinal cohort studies should follow contributors after a stalling event over 6 to 24 months and measure re-entry, subsequent contribution volume, reviewer-author recurrence, and project switching. Such designs would clarify whether stalling typically leads to sustained contributor loss or a recoverable pause. These datasets would also enable stronger causal inference when combined with quasi-experimental methods. For example, researchers could compare contributor outcomes before and after changes in triage policy, reviewer assignment, stale-bot configuration, or reviewer accountability practices using matched comparison projects and differences regarding different style designs. Evidence on recovery trajectories and the effectiveness of post-stalling interventions remains limited in existing PR abandonment research \cite{li2021you, khatoonabadi2023wasted}, and addressing this gap would connect stalling mechanisms to long-term community sustainability outcomes.

\subsubsection{Investigate how stalling costs differ by contributor experience}
Prior work on contributor-abandoned PRs finds that novice contributors and complex PRs are associated with higher abandonment likelihood \cite{khatoonabadi2023wasted}. Our results show that feature-enhancement PRs constitute the largest share of stalled PRs in our dataset, suggesting that a substantial portion of stalled work occurs in contribution contexts that may require broader scope negotiation, design discussion, and sustained coordination. Future research should therefore stratify stalling incidence and post-stalling non-return by contributor tenure, such as first-time, occasional, and regular contributors. Researchers should test whether non-return is distributed uniformly across contributors or concentrated among newcomers. If stalling disproportionately affects newcomers, the community cost compounds because projects lose both the immediate contribution and the pipeline of potential long-term community members. This motivates intervention designs that explicitly protect newcomer trajectories, such as guaranteed first-human-response targets, mentorship escalation, or reviewer assignment for high-scope PRs.

\subsubsection{Evaluate LLM-assisted re-engagement interventions at scale}
This study shows that an LLM-based voting pipeline can feasibly categorize stalled PRs by type at scale. A direct extension is to investigate whether LLMs can support reason-aware re-engagement by generating contextual prompts that summarize the latest blocking event, identify the likely next action, and route the message to the appropriate party, such as the author, reviewer, or maintainer. Such interventions could operationalize cause-aware prevention within existing platform workflows. However, the key research question is not only whether LLM-generated messages improve completion rates, but also whether they do so without increasing noise, perceived pressure, or discussion quality problems. This can be evaluated through randomized controlled field deployments in real open-source projects, measuring contributor return after stalling, PR completion rate, time to next human response, reviewer workload, notification fatigue, and perceived fairness of automated intervention. Future work should also validate LLM-based classifiers against manually labeled gold-standard datasets before deploying them in high-impact workflow interventions.

\subsubsection{Develop ethical and governance guidelines for intervention studies on open-source contributors}
Reason-aware nudges and LLM-generated re-engagement messages introduce human-subjects considerations, including consent expectations, potential coercion or undue pressure, differential effects on newcomers, and risks of public misattribution when automated messages are wrong. Future research should document governance choices explicitly, including opt-out mechanisms, rate limits, transparency about automation, audit procedures, and safeguards against harmful or inaccurate messages. Intervention studies should also consider whether automated prompts place disproportionate pressure on unpaid contributors or create additional emotional burden for maintainers. These concerns are especially important because stalled PRs are public, socially visible interactions, and automated re-engagement may affect contributor reputation or perceived accountability. Researchers should therefore align intervention design with established ethics guidance in mining software repositories and with empirical findings on disruptive bot behaviors \cite{gold2022ethics, wessel2021don, wessel2022bots}.

\section{Threats to Validity}

\subsection{Internal Validity}
One threat concerns the identification of stalled PRs. We collected PRs that were marked as stale or inactive in projects using the Stale bot. While this operationalization is appropriate for studying automation-supported PR inactivity, stale labels may not capture all forms of abandonment. Some PRs may have been administratively closed despite still being viable, while others may have remained inactive without receiving a stale label because of project-specific bot configurations or maintainer practices. To mitigate this threat, we selected projects from a prior study of Stale bot adoption and impact \cite{khatoonabadi2023understanding}, and we interpret our findings as evidence about PRs handled as stale within project workflows rather than all abandoned PRs on GitHub. Another threat concerns PR type classification. Our LLM-based classifier relied on PR titles and descriptions, which may be ambiguous, incomplete, or template-heavy. To reduce this risk, we cleaned PR descriptions, used both titles and descriptions, and adopted a voting-based strategy across multiple LLMs, a common way to improve robustness in classification and annotation settings \cite{dietterich2000ensemble, snow2008cheap}. Cases with three-way disagreement were manually reviewed, and three individuals independently verified a subset of 50 PRs from each category. Nevertheless, some misclassification may remain, especially for PRs that combine multiple intents, such as bug fixing, refactoring, and dependency updates.

The qualitative coding of review comments introduces another potential internal threat because identifying stalling rationales requires human interpretation. Review comments are often short, informal, context-dependent, and embedded in project-specific technical discussions. To mitigate subjective bias, three coders with software development experience participated in the coding process. We used open coding, iterative codebook refinement, double coding, and third-rater adjudication. The resulting average Cohen’s Kappa was $\kappa=0.74$, which indicates substantial agreement according to commonly used interpretation guidelines \cite{cohen1960coefficient, landis1977measurement}. However, we acknowledge that Cohen’s Kappa is only an indicator of agreement and does not eliminate interpretive ambiguity. Some comments may reasonably fit multiple categories, for example, a technically blocked discussion may also involve a delayed reviewer response. We therefore interpret our taxonomy as an empirically grounded abstraction of dominant stalling signals, not as a complete reconstruction of every causal mechanism behind each PR.

\subsection{Construct Validity}
We operationalized stalled PRs through stale-bot-marked PRs in projects where stale automation was part of the contribution workflow. This choice makes the study directly relevant to automation-supported inactivity management, but it also narrows the construct. Stalling, abandonment, and stale closure are related but not identical concepts. A stale PR may still be recoverable, while an abandoned PR may not always be labeled stale. Therefore, our findings should be interpreted as evidence about PRs that became inactive enough to be handled by stale-bot workflows, rather than as a universal measurement of all PR abandonment.

Another construct threat concerns the meaning of ``responsibility'' or ``next required action'' in RQ$_3$. Review comments rarely provide complete evidence about participants' intentions, availability, or private communication. For example, an author may appear unresponsive because the required change is unclear, because they lost interest, or because they discussed the issue outside GitHub. Similarly, a reviewer may appear inactive because of workload, project governance, or because the PR no longer aligns with project priorities. To reduce the risk of blame-oriented interpretation, we frame RQ$_3$ as identifying where the next visible action appears to reside in the GitHub discussion, rather than assigning personal fault. This distinction is important because our data can reveal observable workflow states, but cannot fully reveal participant motivation.

The post-stale engagement measures also involve construct limitations. We measured whether contributors submitted subsequent PRs to the same repository and whether the same reviewers reviewed any of the contributors’ next 10 PRs. These measures capture visible re-engagement within the same project, but they do not capture other forms of continued participation, such as issue discussion, reviewing others’ PRs, contributing to another repository, or communicating through private channels. Thus, a contributor classified as not returning to the same repository may still remain active elsewhere in the open-source ecosystem. Similarly, low reviewer-author recurrence should not be interpreted as direct evidence of interpersonal avoidance or bias. It may also reflect reviewer workload, code ownership, project rotation, or changes in the technical area of later PRs.

\subsection{External Validity}
Our dataset includes 19 popular open-source GitHub repositories that used the Stale bot and were previously examined in related work \cite{khatoonabadi2023understanding}. These projects provide a relevant setting for studying automation-supported PR inactivity, and they span multiple project contexts and programming languages. However, they are not necessarily representative of all open-source projects. Popular repositories often have larger contributor bases, heavier review workloads, more formal contribution practices, and stronger automation infrastructure than smaller or early-stage projects. Therefore, the observed stalling patterns may differ in smaller projects, industrial repositories, projects without stale automation, or communities using platforms other than GitHub.

Another external validity threat is that all projects in our dataset adopted stale-bot workflows. This is appropriate for our research focus, but it may limit generalization to projects that manage inactivity manually or use different automation policies. Stale-bot configuration, inactivity thresholds, exemption rules, and maintainer responses can shape which PRs are labeled stale and how contributors react to those labels. As a result, our findings should be generalized most directly to projects that use similar stale automation practices. Future studies should replicate our analysis across projects without stale bots, across different bot configurations, and across other collaborative development platforms such as GitLab or Gerrit.

The temporal context of the dataset may also affect generalizability. Open-source contribution practices, CI infrastructure, review automation, and LLM-assisted development tools continue to evolve. Consequently, the causes and handling of PR stalling may change as projects adopt more advanced automation, code generation tools, or reviewer recommendation systems. Although our taxonomy captures recurring socio-technical patterns, future replications are needed to determine whether the relative prevalence of communication, technical, and decision-related blockers remains stable over time.

\subsection{Conclusion Validity}
Our conclusions should be interpreted within the scope of our analyses. For RQ$_1$, we identify which PR types are most common among stalled PRs. Although the Chi-square test shows that the distribution of stalled PRs across categories is non-uniform, it does not by itself prove that a category has a higher stalling likelihood than others. Estimating type-specific stalling likelihood would require comparing stalled and non-stalled PRs within each category. For RQ$_2$, our taxonomy is based on sampled general and inline review comment threads. We used random sampling, four independent non-overlapping samples for each comment type, double coding, and third-rater adjudication to improve reliability. However, rare causes may still be underrepresented, and some comments may reasonably fit multiple categories. For RQ$_3$, author return and reviewer re-engagement indicate subsequent visible collaboration patterns, but they should not be interpreted as causal effects of PR stalling.

\section{Conclusion}

This study investigated why pull requests become inactive in pull-based open-source development and how such inactivity affects subsequent collaboration. We analyzed 14,234 stalled pull requests and 164,562 review comments from 19 popular GitHub repositories that used stale-bot workflows. Using an LLM-based voting approach, we classified stalled PRs by contribution type and conducted qualitative coding of general and inline review comments to identify the visible causes of stalling and the location of the next required action. Our findings show that feature-enhancement and issue-fixing PRs constitute the largest share of stalled contributions. Stalling is mainly associated with communication and coordination breakdowns, including missing interaction, delayed feedback, unresolved discussions, and unclear follow-up. Inline comments further show that inactivity is not always caused by disengagement, as many PRs remain stalled because of technical and dependency-related blockers, such as failing checks, configuration problems, compatibility issues, and environment mismatches. The consequences of stalling can also extend beyond a single PR. Only 39.56\% of contributors returned to submit subsequent PRs, and reviewer re-engagement with the same contributors remained limited at approximately 21\%. These patterns suggest that stalled PRs may weaken collaboration continuity and reduce the perceived value of continued participation. Overall, our findings indicate that PR stalling is better understood as a socio-technical coordination problem than as a simple delay or contributor-side abandonment. Time-based stale automation alone is therefore insufficient. Projects need cause-aware interventions that distinguish author inaction, reviewer delay, technical blockage, unresolved decisions, and external dependencies to reduce avoidable contribution waste and support healthier open-source collaboration.

Future work can extend this study by comparing stalled and non-stalled PRs to estimate type-specific stalling likelihood, replicating the analysis across projects without stale bots and across other platforms, and evaluating cause-aware interventions such as targeted re-engagement prompts, CI triage support, and reviewer-assignment mechanisms in real project settings.

\section*{Declarations}
\subsection*{Funding}
This research is supported in part by the Natural Sciences and Engineering Research Council of Canada (NSERC) Discovery Grants program and by the industry-stream NSERC CREATE in Software Analytics Research (SOAR).
\subsection*{Ethical approval}
``Not applicable"
\subsection*{Informed consent}
``Not applicable"
\subsection*{Author Contributions}
Md Shamimur Rahman led the conceptualization and design of the study, collected and preprocessed the data, developed the LLM-based PR classification approach, conducted the empirical analyses, interpreted the findings, and drafted the manuscript. Farhana Akter contributed to the qualitative coding and validation of review comments, helped refine the taxonomy of PR stalling reasons, and reviewed the manuscript. Md Mustakim Billah contributed to the LLM-based classification approach, qualitative coding and validation of review comments, taxonomy refinement, and manuscript review. Zadia Codabux provided methodological guidance, contributed to the interpretation of findings, and critically revised the manuscript. Chanchal K. Roy supervised the overall research, provided strategic direction, contributed to the study design and interpretation of results, and helped finalize the manuscript. All authors reviewed and approved the final version of the manuscript.
\subsection*{Data Availability Statement} Our replication package can be found in the following repository \url{https://doi.org/10.5281/zenodo.20599077}
\subsection*{Conflict of Interest}
We have no conflict of interest.
\subsection*{Clinical Trial Number in the title page}
``Not applicable"
\bibliography{sample-base}

\end{document}